%% file: SmartFlex-EV-Model.tex
\documentclass[preprint,5p]{elsarticle}
 \pdfoutput=1
\usepackage{siunitx}
\usepackage{eurosym}
\usepackage{multirow}
\usepackage{subcaption} 
\usepackage{hyperref} 
\usepackage{booktabs}
\usepackage{float}
\usepackage{array}
\usepackage{amsmath}
\usepackage{amssymb}
\usepackage{enumitem} %
\usepackage{ragged2e}
\usepackage{dblfloatfix}
\usepackage{color}
\usepackage[version=4]{mhchem}

\DeclareSIUnit{\AH}{Ah}
\DeclareSIUnit{\PP}{pp}
\DeclareSIUnit{\sieuro}{\mbox{\euro}}
\DeclareSIUnit{\WH}{Wh}
\DeclareSIUnit{\amperehour}{Ah}
\DeclareSIUnit{\watthour}{Wh}
\DeclareSIUnit{\amperesecond}{As}
\DeclareSIUnit\KWH{kWh}
\DeclareSIUnit{\kmh}{km/h}
\DeclareSIUnit{\Ckm}{\SI{100}{\km}}
\DeclareSIUnit{\kWp}{kW_p}
\DeclareSIUnit\KW{kW}
\DeclareSIUnit{\MWh}{MWh}
\DeclareSIUnit{\TWh}{TWh}
\DeclareSIUnit{\ECT}{\sieuro ct}
\DeclareSIUnit{\million}{\text{million}}
\DeclareSIUnit{\billion}{\text{billion }}

\ExplSyntaxOn
\NewDocumentCommand \hms { o > { \SplitArgument { 2 } { ; } } m }
{
	\group_begin:
	\IfNoValueF {#1}
	{ \keys_set:nn { siunitx } {#1} }
	\siunitx_hms_output:nnn #2
	\group_end:
}
\cs_new_protected:Npn \siunitx_hms_output:nnn #1#2#3
{
	\IfNoValueF {#1}
	{
		\tl_if_blank:nF {#1}
		{
			\SI {#1} { \hour }
			\IfNoValueF {#2} { ~ }
		}
	}
	\IfNoValueF {#2}
	{
		\tl_if_blank:nF {#2}
		{
			\SI {#2} { \minute }
			\IfNoValueF {#3} { ~ }
		}
	}
	\IfNoValueF {#3}
	{ \tl_if_blank:nF {#3} { \SI {#3} { \second } } }
}
\ExplSyntaxOff

\journal{Journal of Applied Energy}

\begin{document}	
	
\begin{frontmatter}
	\title{A Comprehensive Electric Vehicle Model for Vehicle-to-Grid Strategy Development}%
	\author[1,2,3]{Fabian Rücker\corref{cor1}}
	\author[1,2,3]{Ilka Schoeneberger}%
	\author[1,5]{Till Wilmschen}%
	\author[1,2,3]{Ahmed Chahbaz}%
	\author[1,2,3]{Philipp Dechent}%
	\author[1,2,3]{Felix Hildenbrand}%
	\author[1]{Elias Barbers}%
	\author[1,2,3]{Matthias Kuipers}%
	\author[1,2,3]{Jan Figgener}%
	\author[1,2,3,4]{Dirk Uwe Sauer}
	\cortext[cor1]{Corresponding author}
	\address[1]{ Electrochemical Energy Conversion and Storage Systems, Institute for Power Electronics and Electrical Drives (ISEA), RWTH Aachen, Jägerstr. 17-19, 52066 Aachen, Germany; batteries@isea.rwth-aachen.de}
	\address[2]{ Juelich Aachen Research Alliance, JARA-Energy, Germany}
	\address[3]{Institute for Power Generation and Storage Systems (PGS), E.ON ERC, RWTH Aachen University, Germany}
	\address[4]{Helmholtz-Institute Münster (HIMS), Ionics in Energy Storage, Germany}
	\address[5]{Wall Box Chargers, S.L., Spain}
	
	\begin{abstract}
An electric vehicle model is developed to characterize the behavior of the Smart e.d. (2013) while driving, charging and providing vehicle-to-grid services. The battery model is an electro-thermal model with a dual polarization equivalent circuit electrical model coupled with a lumped thermal model with active liquid cooling. The aging trend of the EV’s 50 Ah large format pouch cell with NMC chemistry is evaluated via accelerated aging tests in the laboratory. The EV model is completed with the measurement of the on-board charger efficiency and the charging control behavior via IEC 61851-1. Performance of the model is validated using laboratory pack tests, charging and driving field data. The RMSE of the cell voltage was between \SI{18.49}{\milli \volt} and \SI{67.17}{\milli \volt} per cell for the validation profiles. Cells stored at 100 \% SOC and \SI{40}{\degreeCelsius} reached end-of-life (80 \% of initial capacity) after 431 days to 589 days. The end-of-life for a cell cycled with 80 \% DOD around an SOC of 50 \% is reached after 3634 equivalent full cycles which equates to a driving distance of over \SI{420000}{\kilo \meter}. The full parameter set of the model is provided to serve as a resource for vehicle-to-grid strategy development.

	\end{abstract}
	

\end{frontmatter}
\clearpage

\input{sections/introduction.tex}

\input{sections/methodology_sol_ion_ev_charging.tex}

\input{sections/results.tex}

\bibliography{literatur}
\appendix
\section{Appendix}
\label{sec:appendix}
\begin{table*}[htbp]
	\centering
	\begin{tabular}{ll}
		\toprule
		Abbreviation & Definition \\
		AC & Alternating Current \\
		BMS & Battery Management System \\
		CCS & Combined Charging System \\
		CC & Constant Current \\
		CP & Control Pilot \\
		CV & Constant Voltage \\
		DC & Direct Current \\
		DOD & Depth-of-Discharge \\
		DSO & Distribution System Operator \\
		ECD & Equivalent Circuit Diagram \\
		ECM & Equivalent Circuit Model \\
		EOL & End-of-life \\		
		EQFC & Equivalent full Cycles \\
		EV & Electric Vehicle \\
		EU & European Union\\
		ISEA & Institute for Power Electronics and Electrical Drives \\
		GHG & Greenhouse Gas\\
		LFP & Lithium-Iron-Phosphate \\
		LMO & Lithium-Manganese-Oxide \\
		NCA & Nickel-Cobalt-Aluminum \\	
		NMC & Nickel-Manganese-Cobalt \\
		OCV & Open Circuit Voltage \\
		P2D & Pseudo-two-dimensional \\
		PWM & Pulse-width Modulation \\
		PSO & Particle Swarm Optimization \\
		RMSE & Root Mean Square Error \\
		SECC & Supply Equipment Communication Controller \\
		SEI & Solid Electrolyte Interface \\
		SOC & State of Charge \\
		TIM & Thermal Interface Material \\
		TSO & Transmission System Operator \\
		UDSS & Urban Dynanometer Driving Schedule \\
		V2G & Vehicle-to-Grid\\
		WLTP & World Harmonized Light-duty Vehicle Test Procedure \\ \bottomrule
		\end{tabular}
	\caption{Glossary} 
\end{table*}
\begin{figure*}[htbp]
	\centering
		\includegraphics[width = 0.6\textwidth]{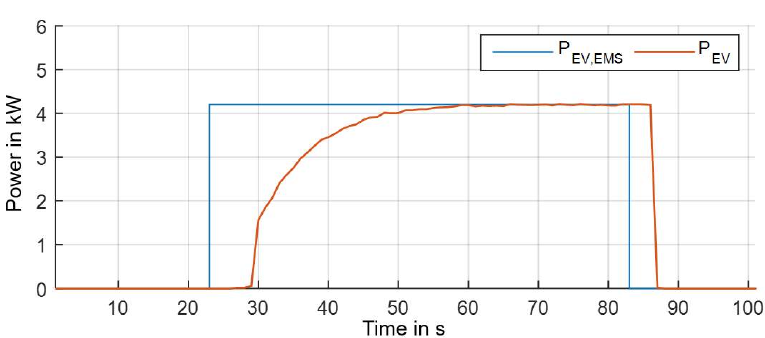}
	\caption{Measurement of charging power control. $P_{EV,EMS}$ is the power requested by the EMS and $P_{EV}$ is the charging power of the EV.}
	\label{fig:emsvsev}
\end{figure*}

\begin{table*}[htbp]
	\centering
	\include{sections/Tables/C1_table}
\caption{Values for parameter $C_1$ in $F$ of electrical model shown in Fig. \ref{fig:eqd-cell} in dependence of cell temperature and SOC.} 
\end{table*}

\begin{table*}[htbp]
		\centering
	\include{sections/Tables/C2_table}
	\caption{Values for parameter $C_2$ in $F$ of electrical model shown in Fig. \ref{fig:eqd-cell} in dependence of cell temperature and SOC.} 
\end{table*}

\begin{table*}[htbp]
		\centering
	\include{sections/Tables/R1_table}
	\caption{Values for parameter $R_1$ in $\Omega$ of electrical model shown in Fig. \ref{fig:eqd-cell} in dependence of cell temperature and SOC.} 
\end{table*}

\begin{table*}[htbp]
		\centering
	\include{sections/Tables/R2_table}
	\caption{Values for parameter $R_2$ in $\Omega$ of electrical model shown in Fig. \ref{fig:eqd-cell} in dependence of cell temperature and SOC.} 
\end{table*}

\begin{table*}[htbp]
		\centering
	\include{sections/Tables/Rser_table}
	\caption{Values for parameter $R_{ser}$ in $\Omega$ of electrical model shown in Fig. \ref{fig:eqd-cell} in dependence of cell temperature and SOC.} 
\end{table*}

\begin{table*}[htbp]
		\centering
	\include{sections/Tables/OCV_table}
	\caption{Values for parameter $OCV$ in $V$ of electrical model shown in Fig. \ref{fig:eqd-cell} in dependence of cell temperature and SOC.} 
\end{table*}

\end{document}

%% file: sections/introduction.tex
\section{Introduction}
The \textit{Transportation} sector is one of the largest contributors to greenhouse gas (GHG) emissions in the world and is the main cause of air pollution in cities. Therefore, many countries and regions around the world have sketched out pathways and adopted regulations in order to reduce GHG emissions of the \textit{Transportation} sector. In the EU, the main elements of the \textit{European Strategy for low-emission mobility} are \textit{Increasing the efficiency of the transport system}, \textit{Speeding up the deployment of low-emission alternative energy for transport} and \textit{Moving towards zero-emission vehicles}. \cite{EuropeanStrategy}
An increased efficiency of the transport system in terms of energy and area use can be achieved with the use of railway, public transport systems and the transformation to cyclist and pedestrian friendly urban areas. Examples for low-emission alternative energy for transport are bio fuels, electricity, renewable synthetic fuels and hydrogen.  In the EU, electricity is a low-emission alternative energy as the share of renewable energy in the electricity sector has increased to 34.1 \% in 2019. \cite{eurostat}
The global number of electric vehicles (EVs) has increased by \SI{400}{\%} from 2016 to 2019 to 4.79 million and is expected to rise in the future. \cite{ieaev} 
The primary use of EVs is transportation and mobility. However, especially privately owned vehicles but also commercial fleet vehicles are only used for mobility for a small portion of the day. As an example, a privately used vehicle in Germany is parked for 97 \% of the time. \cite{infasInstitutfurangewandteSozialwissenschaftGmbH.2017} EVs therefore offer the potential for secondary use as EVs can act as storage systems connected to an electricity grid or a load. Via an internal or external charger, power can be exchanged with the traction battery of the EV.
Several use cases for the secondary use of EVs are being investigated or are already commercially offered. For example, in behind-the-meter use cases the EV can be used as a storage system for on-site energy consumption optimization or an uninterrupted power system (UPS). 
For grid services, EVs can also play an important role. They can offer TSO services such as frequency containment reserve and DSO services such as congestion management and power quality improvement.  The interplay between EVs and renewable energy sources in grids is extensively studied in order to increase the share of renewable energy and avoid grid congestion. Furthermore, EV chargers can be used to form a microgrid by maintaining its voltage and frequency. \newline
For the simulation of the operation of an EV, an EV model is essential. In the case of the simulation of an EV connected to a grid, also the parameterization of the charger and the charging process control is important. This holds especially true for the development and testing of control algorithms for energy-management systems in order to offer aforementioned  services to grid or site operators via vehicle-to-grid (V2G) functionality. 
In addition, the provision of V2G services adds additional loading to the traction battery of the EV. As the traction battery is an EV’s most expensive component, the evaluation of the impact of V2G services on the battery lifetime is important for the economic assessments of such services.
\subsection{Research Objective}
In this paper we parameterize a comprehensive model electric vehicle model for vehicle-to-grid strategy developement for the Smart e.d. (2013). The main contributions of this paper are:
\begin{enumerate}[label=\alph*)]
	\item Electro-thermal model of an EV battery pack
	\item Traction battery break-down (materials, volume and weight distribution)
	\item Accelerated aging tests of EV cell (cycle and calendar tests)
	\item Efficiency measurements of the on-board charger
	\item Parameterization of the control of the charging process according to IEC 61851-1
	\end{enumerate}
\begin{table*}[htbp]
	\footnotesize
	\begin{tabular}{m{0.08\textwidth}>{\centering} m{0.03\textwidth}>{\raggedright} m{0.50\textwidth}>{\centering} m{0.05\textwidth}>{\centering} m{0.05\textwidth}>{\centering} m{0.05\textwidth}>{\centering\arraybackslash} m{0.1\textwidth}}
		\toprule
		Source & Date & Focus and Results & \multicolumn{3}{c}{Model} & Level \\ 
		 & & & Electrical & Thermal & Aging & \\ \hline
		 Meshabi et al. \cite{MESBAHI2021102260} & 2021 & - Electro-thermal model for a \ce{LiNiMnCoO2} pouch cell with distributed dual polarization (2RC) electrical circuits and distributed thermal (RC) circuits. \newline - Relative error of less than \SI{1.35}{\degreeCelsius} was achieved in a constant current discharge profile with a 1C rate. & \checkmark & \checkmark & x & Cell \\ \hline
		 Schmid et al. \cite{SCHMID2020101736} & 2020 & - Matrix-vector-based framework for modeling and simulation of EV battery packs with \ce{LiNiMnCoO2} automotive cells\newline - Dual polarization electrical model (2RC)  \newline - Modified Cauer thermal model for each cell in a battery pack and heat transfer between cells, contacts and bus bars of the pack \newline - Holistic aging model for calendar and cycle aging \newline - Model allows for the investigation of three fault cases in the battery pack: Increased contact resistance, external short circuit, internal short circuit  & \checkmark & \checkmark & \checkmark & Pack \& Cell \\ \hline
		 Zhu et al. \cite{ZHU2019113339} & 2019 & - Dual polarization electrical model (2RC) of a \ce{LiNiMnCoO2} cell \newline - Excitation of the battery by inverse binary sequences which eliminates drift of operating conditions and even-order non-linearities \newline - Parameter identification by particle swarm optimization (PSO). \newline The RMSE under the urban dynanometer driving schedule (UDSS) of the terminal voltage was \SI{8.61}{\milli \volt}. & \checkmark & x & x & Cell \\ \hline
		 Wen et al. \cite{electronics8080834} & 2019 & - Dual polarization electrical model (2RC) of a \ce{LiNiMnCoO2} cell \newline - Parameter identification via recursive least square method with data from pseudo random binary sequence tests \newline - Improved precision for parameters via stochastic theory response reconstruction in contrast to the use of a butterworth filter. & \checkmark & x & x & Cell \\ \hline
		 Irima et al. \cite{8892901} & 2019 & - EV model of a Renault Zoe consisting of the following parts: Vehicle, Driver, Vehicle Control Unit, Electric Motor and Battery.\newline - Two electrical equivalent models: RC model and 3RC Thevenin model. \newline - Simulation of a speed profile with the Simcenter Amesim platform. & \checkmark & x & x & Pack \& Cell  \\ \hline
		 Hosseinzadeh et al. \cite{HOSSEINZADEH201877} & 2018 & - 1D electrochemical-thermal model of a \ce{LiNiMnCoO2} pouch cell for an EV. \newline - 3D lumped thermal model of cell. \newline - Decrease of ambient temperature from \SI{45}{\degreeCelsius} to \SI{5}{\degreeCelsius} leads to a capacity drop by 17.1 \% for a 0.5C discharge and a power loss of 7.57 \% under WLTP drive cycle. & \checkmark & \checkmark & x & Cell \\ \hline
		 Jafari et al. \cite{8070984} & 2018 & - EV battery cycle aging evaluation for driving and vehicle-to-grid services\newline - Cycle aging model for \ce{LiFePO4} cells with dependency on C-rate, total Ah throughput and temperature& x & x & \checkmark & Cell \\ \hline
		Gao et al. \cite{GAO2017103} & 2017 & - Measurement of capacity degradation and resistance increase for \ce{LiCoO2} 18650 cells\newline - Aging mechanisms are identified via incremental capacity analysis: Loss of active material and loss of lithium inventory. \newline - Overall aging accelerates dramatically for rates over 1C and when the cut-off voltage exceeds \SI{4.2}{\volt}. \newline - Establishment of a capacity degradation model. & x & x & \checkmark & Cell \\ \hline
		Jaguemont et al. \cite{7006731} & 2016 & - Modeling of a Hybrid-EV Lithium-Ion (\ce{LiFePO4}) battery pack at low temperatures \newline - 2-D battery pack electro-thermal model applicable to subzero temperatures \newline-  Thevenin electrical model and ohmic losses, conductive and convective heat transfer in thermal model & \checkmark & \checkmark & x & Pack \& Cell \\ \hline
		Schmalstieg et al. \cite{Schmalstieg.2014} & 2014 & - Holistic aging model for \ce{LiNiMnCoO2} based 18650 lithium-ion cells \newline - Calendar aging tests for different storage SOCs and temperatures and cycle aging tests for different DODs and average SOCs. \newline - Capacity and inner resistance trend measured with a 1C discharge and pulse power characterization profile respectively.\newline - Electric model consists of series resistance, two ZARC elements and an OCV source parameterized by EIS measurements. & \checkmark & \checkmark & \checkmark & Cell \\ \hline
		\bottomrule		
	\end{tabular}
	\caption{Summary of selected literature about li-ion battery and EV modeling.}
	\label{tab:batterymodel}
\end{table*}
\subsection{Literature Review}
The relevant parts of an EV model for V2G applications are the \textit{battery model}, the \textit{charger model} and the \textit{charging control model}. \newline 
\textit{Battery models} in literature have been mainly divided into three categories for the electrical component: 
Physics-based electrochemical models, equivalent circuit models and data-driven models. \cite{WANG2020110015} In table \ref{tab:batterymodel} selected literature about li-ion battery and EV modeling is summarized with regards to their focus, results and modeled components. \newline
Physics-based electrochemical models trace back to the work of Newman and Tiedermann \cite{Newman1975PorouselectrodeTW} and were extended by Fuller \cite{osti_142201} for lithium-ion batteries with intercalation. An extensive review of the electrochemical processes in a battery can be found in \cite{electrochemicalsystems}.
In a single-particle model, a radial diffusion equation describes the lithium-ion diffusion in the solid phase of one representative particle for each electrode. \cite{ROMEROBECERRIL201110267} In pseudo-two-dimensional (P2D) models each electrode is composed of several spherical particles and the impact of the electrolyte is taken into account. Numerous partial differential equations describe the reactions inside the cell which leads to a large number of unknown variables that need to be identified using global optimization methods. \cite{WANG2020110015} \newline
In electrical equivalent circuit models an electrical circuit is proposed and its components are parameterized through measurements such as impedance spectroscopy, pulse tests and open-circuit voltage (OCV) measurements. In literature, two types are used: integral-order models and fractional-order models. Equivalent circuits can vary in their number and type of components which has an impact on the accuracy and computational complexity of the model. The simplest model is the Rint model that consists of an ideal voltage source in series with a resistor. \cite{JOHNSON2002321} In order to account for transient processes with different time constants such as the charge-transfer and diffusion phenomena, RC networks can be utilized. In \cite{6652363, Liaw2004835,6237284,6108373} the Rint model is extended with one RC element. \newline
Other studies use data-driven methods, i.e. machine learning, to parameterize battery models. \cite{MINGANT2021102592,TANG2019113591,TANG2021100302}
Further studies also model the hysteresis behavior of the OCV of lithium-ion batteries as was done in \cite{Carlier2002} for \ce{LiCoO2} cells and in \cite{TRAN2020101785} for \ce{LiFePO4} cells.
In the study conducted by Tran et al. the hysteresis effect was stronger in lithium-ion batteries with LFP and NCA chemistry compared to NMC and LMO chemistry. \cite{batteries7030051}
Electrical battery models are coupled with thermal models as the electrical parameters, such as the inner resistance, are temperature dependent.
An example for the coupling of a thermal 3D model with a P2D model can be found in \cite{Habedank_2018}. Yang et al. employ machine based learning to the thermal parameterization of EV Li-Ion batteries from external short circuit experiments. \cite{YANG2020}

In addition to an electro-thermal battery model, an aging model of the traction battery is relevant for EV simulation models.
Previous cycle and calendar aging tests have been conducted in \cite{ECKER2012248,ECKER2014839} and the aging trend in aging tests were evaluated with periodic check ups which include a capacity test (full discharge), impedance spectroscopy and pulse tests. In \cite{LEWERENZ201757} the authors used differential voltage analysis in order to evaluate calendar and cycle aging of a \ce{LiFePO4} cell. Further extensions of a battery model treat mechanical stress during charging and discharging \cite{Renganathan_2010,LARESGOITI2015112} or lithium-plating \cite{osti_20001062}. \\
Within the AVTE project, conducted in the US, numerous EVs were operated and extensively tested. Among others, also the Smart e.d. was tested. The researchers conducted battery tests, such as static capacity tests and pulse power characterization tests along the lifetime of the EV. After two years of operation and \SI{19000}{\kilo \meter} the traction battery of the Smart e.d. lost \SI{6.6}{\percent} of its capacity and \SI{15.9}{\percent} of its \SI{30}{\second} discharge power capability at \SI{80}{\percent} depth-of-discharge (DOD).\cite{INLvehicletesting}
For an EV model for V2G applications a \textit{charger model} and a \textit{charging control model} are essential components.
 In \cite{7556282}, the authors developed an on-board charger prototype that achieves a peak efficiency value of \SI{97.3}{\percent} in boost operation mode and \SI{97}{\percent} in buck operation mode. The on-board charger developed by Radimov et al. is a bi-directional, three-stage, on-board charger with a peak efficiency of \SI{96.65}{\percent}. \cite{radimov2020three} Schram et al. determined the V2G round-trip efficiency of a Renault Zoe with a bi-directional on-board charger to be \SI{85.1}{\percent} and of a Nissan Leaf connected to an external charging station to be \SI{87.0}{\percent}. \cite{9203459}
In the Parker project, grid services were offered with a V2G setup using commercial DC-chargers and commercial EVs using CHAdeMO DC-charging. The researchers set power set-points and evaluated that the provided power by the charger lagged \SI{7}{\second} behind the requested power and the set-point error was \SI{8.7}{\percent}. The maximum charger efficiency of the \SI{10}{\KW} chargers was \SI{86}{\percent} and the efficiency exhibited a large drop at charging power below \SI{20}{\percent} of the rated power. \cite{PeterBachAndersenSeyedmostafaHashemiToghroljerdiThomasMeierSrensenBjrnEskeChristense.2019} Another project that investigated V2G services with EVs was the INEES project in Germany. In this project, experimental \SI{10}{\KW} DC charging stations were used with VW eUps that use a CCS plug system. The power set-point was reached almost instantaneously with this setup. \cite{INEES} In the provision of power by a fleet of EVs it was observed that the power set-point for the fleet was reached within \SI{1}{\second}. \cite{degner2017grid}

%% file: sections/methodology_sol_ion_ev_charging.tex
\section{Methodology}

We parameterize and validate the EV model with the Smart electric drive (3rd Generation \cite{DaimlerAG}, production year and manufacturer: 2013, Daimler AG). The specifications of the Smart e.d. (3rd Generation) are summarized in Table \ref{tab:evspecs}. The model is implemented in Matlab\textsuperscript \textregistered/Simulink.

\begin{table}[h]
	\centering
	\begin{tabular}{ l l}
		\toprule
		\multicolumn{2}{l}{\textbf{General Specifications}} \\ 
		Max. speed & \SI{125}{\kmh}  \\  
		Acceleration 0-\SI{100}{\kmh}  & \SI{11.5}{\second}  \\  
		Weight  & \SI{900}{\kilogram}  \\ 
		\multicolumn{2}{l}{\textbf{Traction Battery}} \\ 
		Chemistry & Li-Ion (NMC/Graphite)  \\
		Nominal Capacity & \SI{17.6}{\KWH} \\
		Rated/Max Voltage & \SI{339}{\volt}/\SI{391}{\volt} \\
		Rated capacity & \SI{52}{\AH} \\ 
		Layout & 93s1p  \\
		Weight & \SI{179.6}{\kilogram} \\ 
		Permissible Temperature & \SI{-25}{\degreeCelsius} - \SI{+55}{\degreeCelsius} \\
		\multicolumn{2}{l}{\textbf{Electric Motor}} \\
		Motor type & AC synchronous motor \\   
		Max. output & \SI{55}{\KW}\\
		Max. continuous output & \SI{35}{\KW} \\
		Peak Torque & \SI{130}{Nm}\\
		Max. rpm & \SI{11800}{}\\
		\multicolumn{2}{l}{\textbf{On-board Charger}} \\
		Standard & IEC62196-2 \& ISO 155118 \\
		Type & 1-phase AC \& 3-phase AC \\
		Max. Power & \SI{22}{\KW} \\ \bottomrule \hline
	\end{tabular}
	\caption{Specifications of Smart e.d. (3rd Generation). \cite{introductionsmart}}
	\label{tab:evspecs}
\end{table}
\subsection{Electric vehicle model}
The EV model is divided into several parts:
\begin{enumerate}
	\item Traction battery model 
	\begin{enumerate}
		\item General  (Section \ref{sec:batterymodel})
		\item Electrical model (Section \ref{sec:electricalmodel} and \ref{sec:resultselectricalmodel}))
		\item Thermal model and traction battery pack materials, volume and weight distributions (Section \ref{sec:thermalmodel} and \ref{sec:resultsthermalmodel})
		\item Aging model (Section \ref{sec:agingmodel} and \ref{sec:resultsagingmodel})
	\end{enumerate}
	\item BMS and charger model (Section \ref{sec:chargerbms} and \ref{sec:chargerresults})
	\begin{enumerate}
		\item Charger efficiency	
		\item Charging control model 
	\end{enumerate}
\end{enumerate}

\subsubsection{Traction Battery}
\label{sec:batterymodel}
\begin{figure*}[t]
	\begin{minipage}[b]{0.48\textwidth}
		\centering
		\subcaptionbox{Smart e.d. (2013) in front of a charging station.\label{fig:smartev}}{\includegraphics[width=\textwidth]{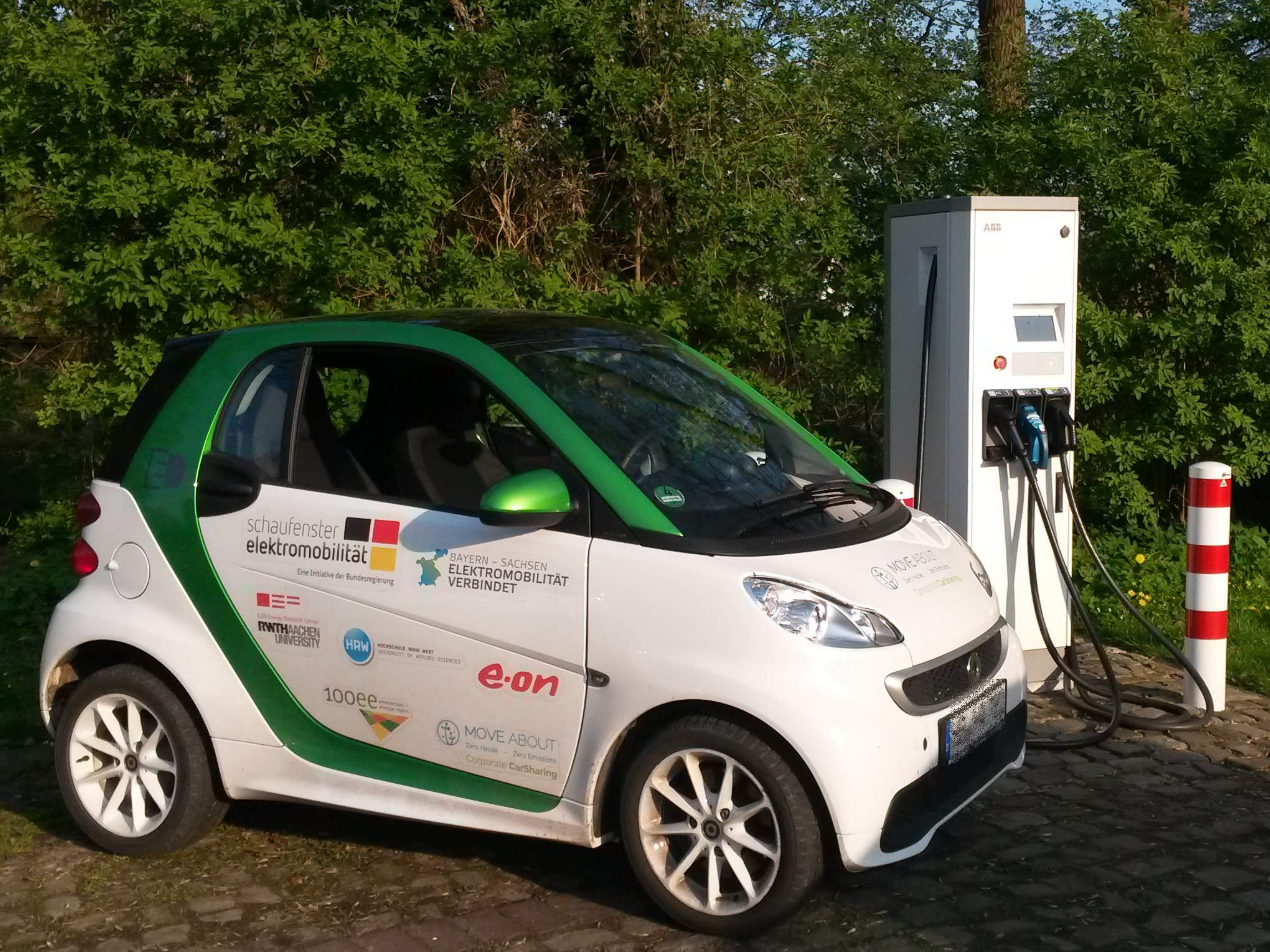}}
	\end{minipage}
	\begin{minipage}{0.02\textwidth}
		\hfill
	\end{minipage}
	\begin{minipage}[b]{0.48\textwidth}
		\centering
		\subcaptionbox{Single Module with outer plate taken off with a visible single cell of the 31 cells of the module.\label{fig:singlemodule}}{\includegraphics[width=\textwidth]{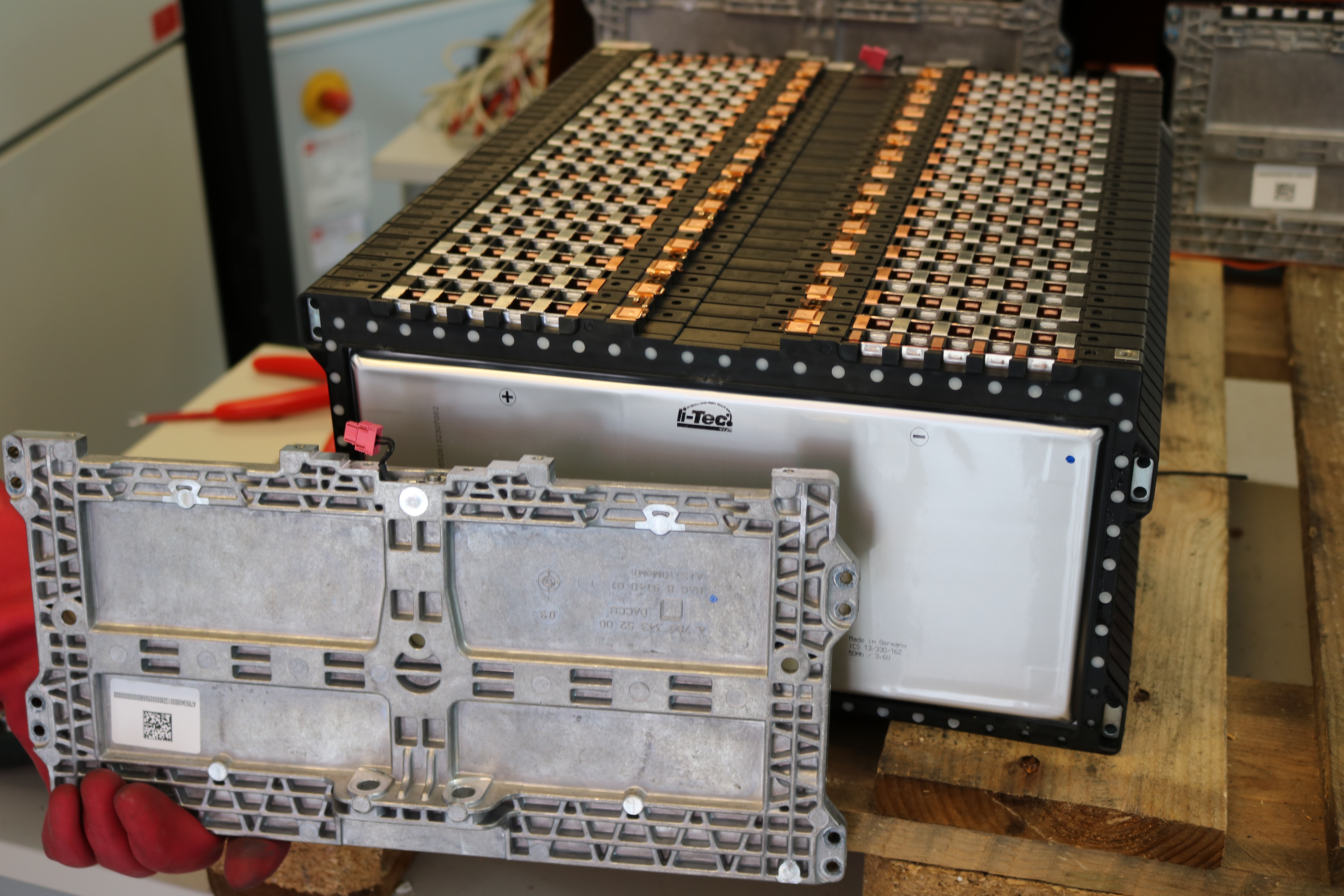}}
	\end{minipage}
	\begin{minipage}[b]{0.48\textwidth}
		\centering
		\subcaptionbox{Traction battery pack of Smart e.d.\label{fig:packwithcase}}{\includegraphics[width=\textwidth]{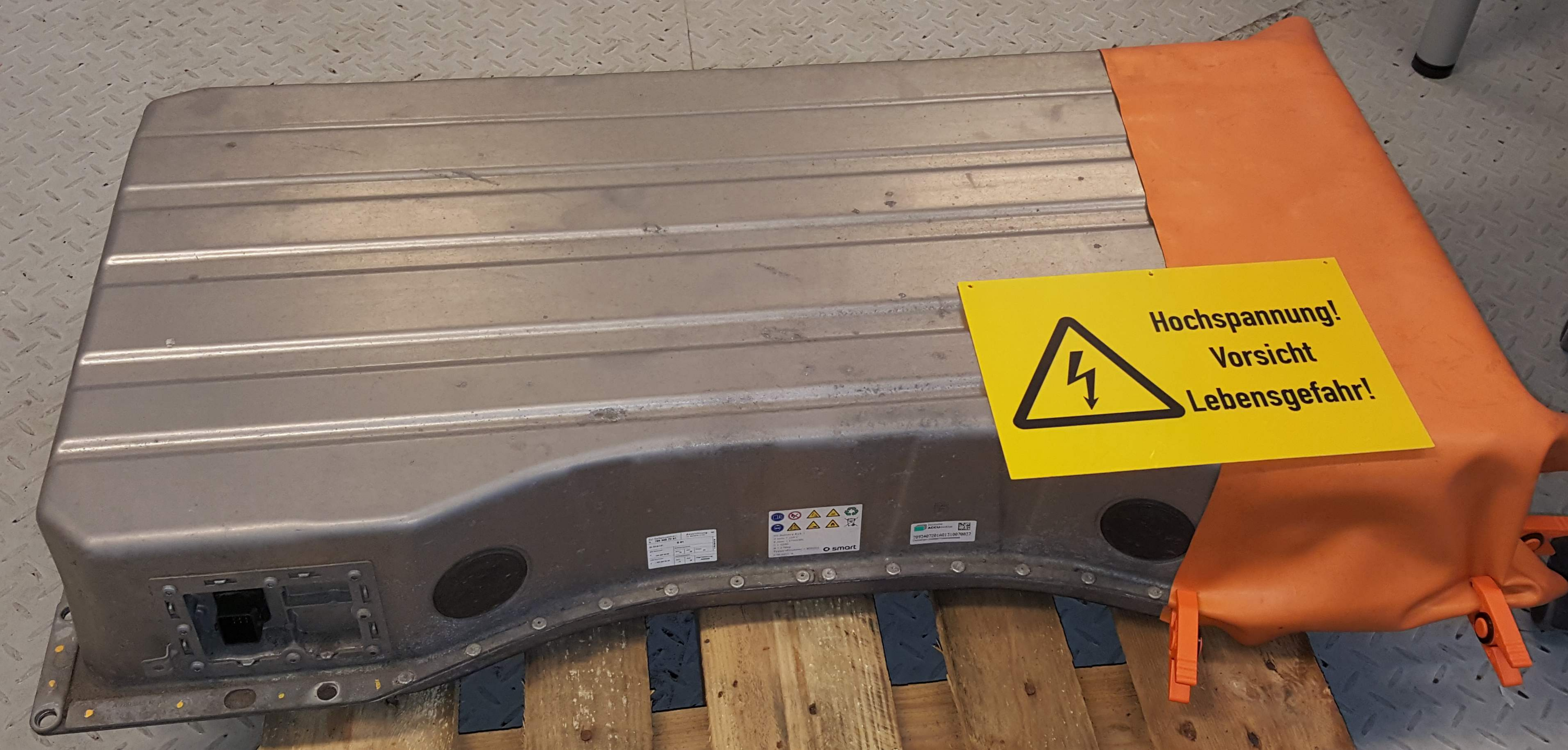}}
	\end{minipage}
	\begin{minipage}{0.02\textwidth}
		\hfill
	\end{minipage}	
	\begin{minipage}[b]{0.48\textwidth}
		\centering
		\subcaptionbox{Traction battery pack of Smart e.d. without casing. 3 Modules with 31 cells each connected in series (93s1p) with liquid cooling plates and slave BMS. Bottom left:Master BMS. Bottom right: DC connector.\label{fig:packwithoutcase}}{\includegraphics[width=\textwidth]{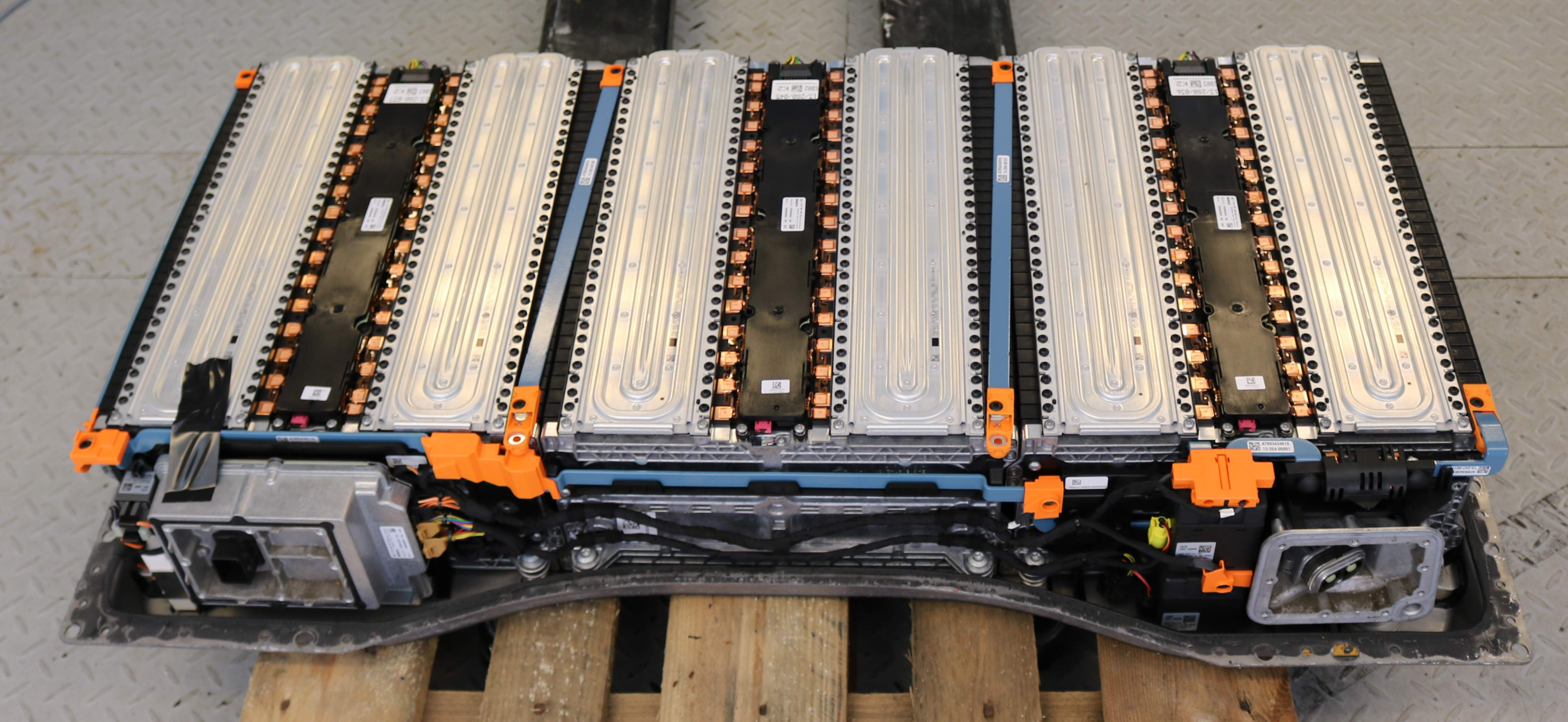}}
	\end{minipage}
	\caption{Pictures of the Smart e.d., the battery pack and a single battery pack module.}
\end{figure*}

The traction battery of the Smart electric drive (2013) is mounted at the bottom of the vehicle, has a battery layout of 93s1p and a capacity of \SI{17.6}{\KWH}. The  battery management system (BMS) limits the the usable SOC range to 3.2\% - 95.3\%, which results in a usable battery capacity of \SI{16.2}{\KWH}. More specifications about the battery given by the official data sheet can be found in table \ref{tab:evspecs}. The traction battery is housed in a case (see Fig. \ref{fig:packwithcase}) that is constructed from a bottom part made of steel and a top part made of aluminum. The battery consists of 3 modules with 31 cells each that are all connected in series  (see Fig. \ref{fig:packwithoutcase} and Fig. \ref{fig:singlemodule}). In addition to the modules, the pack contains the Master BMS, the DC connector, HV contactors, shunt, precharge circuit, cooling system pipes and a desiccant cartridge. \\
Each module houses 31 cells which are held in place by plastic frames and are electrically connected via copper connectors. Two metal parts at each end and metal rods that go through the whole module provide stability. The slave BMS sits between the aluminum cooling plates on top of the module. The cooling plates cool (or heat) the terminals of the cells which allows cooling (heating) within the cell. Between the copper conductor and the cooling plate, strips of thermal interface material (TIM) are placed such that the aluminum cooling plates do not short-circuit the cells. Within the cooling plates a coolant (water/glycol mixture) circulates in order to cool the battery pack during driving and charging. If the battery temperature falls below \SI{0}{\degreeCelsius} the coolant is heated in order to ensure safe operation of the battery pack  \cite{introductionsmart}.

The battery cell is the \textit{HEA50 high energy cell} (ICS 13/330/162, IMP 13/330/162) manufactured by Li-Tec (Daimler) with a nominal capacity of \SI{52}{\AH}. 
The manufacturer also provides details about the aging characteristics of the cell in the data sheet (see table \ref{tab:cellspecs}). Cycle life time is given as 3000 cycles at 100\% depth of discharge (DOD) and a charge/discharge rate of $2C/2C$. The calendaric lifetime is given as $\geq 5 $ years of shelf life at \SI{50}{\percent} SOC and \SI{-30}{\celsius} - \SI{25}{\celsius}. 

\begin{table}[h]
	\centering
	\begin{tabular}{m{0.35\columnwidth} m{0.6\columnwidth}}
		\toprule
		\multicolumn{2}{l}{\textbf{Battery Cell Specifications:}} \\
		\multicolumn{2}{l}{\textbf{HEA50 high energy cell}} \\
		\multicolumn{2}{l}{\textbf{(ICS 13/330/162, IMP 13/330/162)}} \\ 
		Manufacturer  & li-Tec (Daimler AG)  \\  
		SOC operation limit   &  \SI{3.2}{\percent} - \SI{95.3}{\percent} \\  
		Nominal capacity  & \SI{52}{\AH} (\SI{0.5}{C} discharge)  \\
		Nominal voltage & \SI{3.65}{\volt} \\
		Voltage range & \SI{3.0}{\volt} - \SI{4.2}{\volt} \\
		Temperature range & \SI{-25}{\degreeCelsius} - \SI{+55}{\degreeCelsius} \\
		Gravimetric energy density & \SI[per-mode=symbol]{147}{\WH \per \kilogram} \\
		Volumetric energy density & \SI[per-mode=symbol]{280}{\WH \per \liter} \\
		Inner resistance &  $\leq$ \SI{1.8}{\milli \ohm} (\SI{5}{\second}, \SI{200}{\ampere}, \SI{50}{\percent} SOC) \\
		\multirow{2}{*}{Cathode} & Li-Ion with LITARION \textsuperscript{\textregistered} NMC\\ 
		&  (33 \% Ni, 33 \% Co, 33 \% Mn \cite{Witzenhausen:687819}) \\
		Anode & Graphite \\
		Anode terminal & Copper \\
		Cathode terminal & Aluminum \\
		Separator & Ceramic (SEPARION\textsuperscript{\textregistered}) \\
		Cell case material  & PET \\
		Width x length x depth & \SI{32.8}{\centi \metre} x \SI{16.1}{\centi\metre} x \SI{1.3}{\centi\metre} (\SI{50}{\percent} SOC) \\
		Weight & \SI{1296.5}{\gram} \\ \bottomrule \hline
	\end{tabular}
	\caption{Specifications of battery cell HEA50 high energy cell (ICS 13/330/162, IMP 13/330/162) of Smart e.d. (3rd Generation). \cite{LiTecBatteryGmbH.2015}}
	\label{tab:cellspecs}
\end{table}

\subsubsection{Electrical Model}
\label{sec:electricalmodel}
\begin{figure}[H]
	\centering
	\includegraphics[]{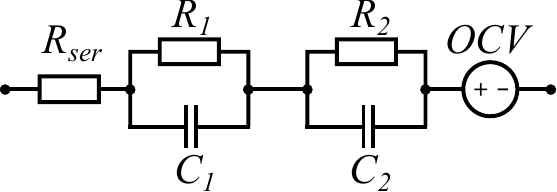}
	\caption{Equivalent circuit diagram of battery cell.}
	\label{fig:eqd-cell}
\end{figure}  
The electrical model of the traction battery is based on the model of a single battery cell. Cell-to-cell variations within the pack in terms of capacity and inner resistance are neglected. Hence, also a balancing system is not required. This simplification leads to faster simulation times and is referred to in literature as a model on pack level  in contrast to models on material or cell level \cite{Abada.2016}. \\
In systems theory, classification of models are based on their physical interpretability. With this in mind, battery models can be divided into three categories: white box models, grey box models and black box models \cite{JanPhilippSchmidt.2013}. In this work a grey box model based on an equivalent circuit model (ECM) is used to model the electric behavior of the battery cell. A multitude of different ECMs is used in literature, such as in Ref. \cite{Hu.2012}, which differ in their complexity, accuracy and required computational power. The ECM of choice in this work is shown in Fig. \ref{fig:eqd-cell} which is expected to provide a good trade-off between computational requirements and accuracy for the simulations. It consists of a voltage source, representing the OCV, in series with a resistance $R_{ser}$ and two RC-elements. It is referred to in literature as a dual polarization model (Thevenin 2RC).\\
All parameters are dependent on the state of charge (SOC) and temperature of the cell. OCV measurements were carried out with a battery cell in a climate chamber while regulating the temperature. During the measurement the cell was discharged. For each SOC state the cell voltage was measured after a relaxation period (period with no load on the cell) such that the measured voltage can be regarded as the OCV. Additionally, two RC-elements and an ohmic resistance $R_{ser}$ were parameterized  in order to model physical processes occurring within the cell that have an impact on the electrical behavior. The resistance $R_{ser}$ is the ohmic resistance due to limited conductivity for electrons and ions within the cell. It leads to instantaneous voltage drops when the cell is under load. The first RC element models the intercalation/de-intercalation of Li-Ions into/from the electrodes where $R_1$ models the charge transfer resistance and $C_1$ models the double layer capactity at the respective electrodes. The second RC element models the concentration gradient of Li-Ions and diffusion in the electrolyte \cite{Schmalstieg.2017,8398150,9236517}.\\The dynamic processes within the cell have different speeds. The overvoltage at a reaction surface with double layer capacitance builds up within milliseconds, the inhomogeneous  electrolyte concentration reaches a steady state after one or several minutes and the solid state diffusion overvoltage builds up and decays even more slowly. \cite{Witzenhausen:687819} \\ The resistance and capacity parameters of the equivalent circuit diagram for the cell were fitted with impedance spectroscopy measurements using a Digatron EIS-Meter in a frequency range of \SI{1}{\milli  \hertz} to \SI{6}{\kilo \hertz} and a temperature regulated climate chamber. Further information about impedance spectroscopy measurements with EIS-Meters can be found in the dissertation of Kiel \cite{Kiel:228547}. Further information about extraction of ECM parameters and their interpretation with regards to physical processes within the cell can be found in the dissertation of Witzenhausen \cite{Witzenhausen:687819}. Impedance spectra at different SOCs and temperatures of the cell were measured and the parameters were extracted from resulting Nyquist diagrams. The results are shown in Fig. \ref{fig:elparameters} and described in section \ref{sec:resultselectricalmodel}.

\subsubsection{Thermal Model}
\label{sec:thermalmodel}
\begin{figure}[H]
	\centering
	\includegraphics[]{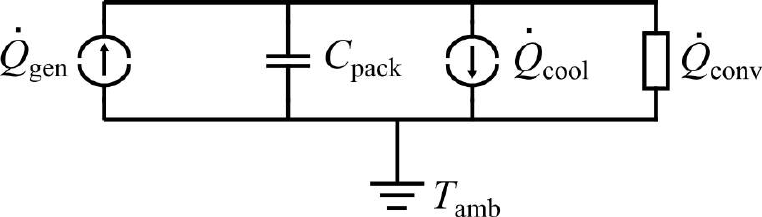}
	\caption{Equivalent circuit model of the thermal model of the battery pack.}
	\label{fig:eqc:thermal}
\end{figure}

The temperature of the battery pack has a direct influence on its electrical performance, capacity, efficiency and safety. During the operation of the battery pack, heat is produced. According to Bernardi et al. \cite{Bernardi_1985} the following equation describes the heat generation current $\dot{Q}$ in a cell
\begin{equation}
	\dot{Q}_{gen} = \dot{Q}_{irr}  +\dot{Q}_{rev} +\dot{Q}_{react} + \dot{Q}_{mix},
\end{equation}
where 
\begin{itemize}
	\item $\dot{Q}_{irr}$ is the irreversible ohmic heat generation,
	\item $\dot{Q}_{rev}$ is the reversible heat generation due to the intercalation and deintercalation of ions at the electrodes,
	\item $\dot{Q}_{react}$ is the heat generation due to side reactions of the electrolyte with the electrodes (i.e., phase changes) and
	\item $\dot{Q}_{mix}$ is the heat generation associated to the relaxation of concentration profiles.
\end{itemize}
The heat currents $\dot{Q}_{react}$ and $\dot{Q}_{mix}$ are often neglected in literature in lithium-ion battery modeling \cite{JanPhilippSchmidt.2013}.
In this model also only $\dot{Q}_{rev}$ is not considered. This simplification can also be found in the work of Magnor  \cite{Magnor:696065}. In order to model the reversible heat, the entropy coefficients of the cell would have to be determined. In theory, the parameterization of the entropy coefficients could be achieved with the measurement of the voltage response after a temperature change of the cell. The conducted OCV measurements however, did not provide the sufficient accuracy in order to accurately parameterize the entropy coefficients. As the irreversible ohmic heat generation $\dot{Q}_{irr}$ is dominant for large currents \cite{Hust:752755} and a lumped thermal model is used in this publication, we deem this to be an acceptable simplification. \\
In order to model the thermal behavior of a battery pack a simplified reduced order model is used. This is done to reduce the numerical solving effort and computation time for long simulation time frames. The battery pack of the Smart e.d. is cooled during driving and charging by a liquid cooling system in contrast to other battery packs which are cooled by forced convection with the use of a Fan (i.e., Mitsubishi iMiEV) or free convection (i.e., VW eUp). Furthermore, privately owned vehicles only spend less than $\SI{4}{\percent}$ of their time driving  (own analysis with data of \cite{mobility95} in Germany). The use of a simplified reduced order model is therefore assumed to be sufficient to account for the impact of temperature on performance and aging of the battery pack. \\
The equivalent circuit model for the reduced order model is shown in Fig. \ref{fig:eqc:thermal}.
The thermal model equation is then
\begin{align}
	C_{pack} \cdot  \frac{dT_{pack}}{dt} = \dot{Q}_{gen}-\dot{Q}_{diss},
\end{align}
with
\begin{equation}
	C_{pack} = m_{pack} \cdot C_{pack-spec}.
\end{equation}
Here, $C_{pack}$ is the specific heat capacity of the battery pack, $m_{pack}$ is the battery pack mass, $\dot{Q}_{gen}$ and  $\dot{Q}_{diss}$ are the heat generation rate and heat dissipation rate respectively. 
The dissipation rate $\dot{Q}_{diss}$  is seperated into two components
\begin{equation}
	\dot{Q}_{diss} = \dot{Q}_{cool} + \dot{Q}_{conv},
\end{equation}
where $\dot{Q}_{conv}$ is the cooling rate due to the heat transfer to the surroundings by convection and $\dot{Q}_{cool}$ is the cooling/heating rate due to the the liquid cooling system.
The convection rate is calculated as 
\begin{equation}
	\dot{Q}_{conv} = (\alpha_x + \alpha_y + \alpha_z) \cdot \Delta T
\end{equation}
with $ \Delta T = T_{pack} - T_{ambient}$ and $\alpha$ being the convection coefficient in one spatial direction:
\begin{equation}
	\alpha = \alpha_{spec} \cdot A
\end{equation}
$A$ is the surface area and $\alpha_{spec}$ is the specific heat transfer coefficient. \\
The cooling system of the drive train components cools the traction battery, charger, engine, power electronics control unit and motor. It consists of two coolant pumps, traction battery heater, chiller, electric expansion valve and an electromotive water valve. The coolant is a glycol/water (50:50) mixture. The BMS can also decouple the coolant circuit of the traction battery from the rest of the cooling circuit via the electromotive water valve if the traction battery needs specific cooling. \cite{introductionsmart}\\ 
The calculation of the flow rate and cooling power of the liquid cooling system is based on the model done by Cédric \cite{AngeEtienneAcquaviva.2012}. The flow rate depends linearly on the battery temperature and the coolant temperature is assumed to be equal to the ambient temperature. The cooling power provided by the liquid cooling system is
\begin{equation}
	\dot{Q}_{cool} = \Delta T \cdot \rho_{coolant} \cdot c_{coolant} \cdot v_{coolant},
\end{equation}
with the density $\rho_{coolant} = \SI{1080}{\kilogram \per \meter^3}$ , specific heat capacity $c_{coolant} = \SI{3320}{\joule \per \kilogram \per \kelvin}$ and 
\begin{equation}
	v_{coolant} = \left|\frac{T_{pack}}{45}\cdot \SI[per-mode=symbol]{1.513e-5}{\meter^3 \per (\second \degreeCelsius)}\right|
	\label{eq:flowrate}
\end{equation}
which is devised from information found in \cite{AngeEtienneAcquaviva.2012}.\\ 
At low temperatures the inner resistance of the battery increases due to, among other effects, lower ion conductivity of the electrolyte. When charging a lithium-ion battery at low temperatures, high surface area lithium deposition on the graphite anode, also known as plating, can occur, which is a safety issue due to short-circuit risks \cite{FRIESEN20161}. \\
Therefore, during the charging process the coolant is heated when the battery temperature falls below \SI{0}{\degreeCelsius} \cite{introductionsmart}. To take this into account, during charging, the temperature of the battery pack is kept above $\SI{0}{\degreeCelsius}$ in the model.

\begin{table}[h]
	\begin{tabular}{cc}
		\toprule
		\textbf{Material} & \textbf{Specific heat capacity in $\SI[per-mode=symbol]{}{\kilo \joule \per \kilogram \per \kelvin}$} \\ 
		Cell         & 1095 \cite{C.Ziebert}                                                       \\ 
		Steel        & 502 \cite{Baehr}                                                            \\ 
		Aluminum     & 891 \cite{DOWNIE1980779}                                                    \\ 
		Plastic (PP) & 1570 \cite{WEIDENFELLER2004423}                                             \\ 
		Air          & 1.01 \cite{C.Ziebert}                                                       \\ \bottomrule \hline
	\end{tabular}
	\caption{Literature values for specific heat capacity}
	\label{tab:thermalspecs}
\end{table}

\subsubsection{Aging tests}
\label{sec:agingmodel}
\begin{figure}[H]
		\centering
		\includegraphics[width = \columnwidth]{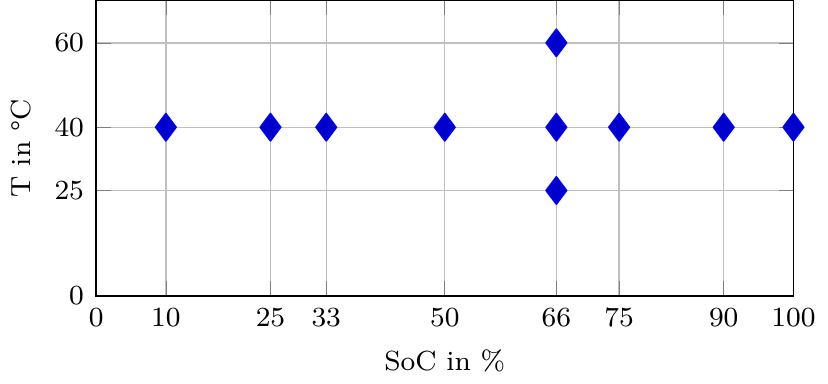}
		\caption{Calendar aging test conditions.}
		\label{fig:caltestconditions}
		\includegraphics[width = \columnwidth]{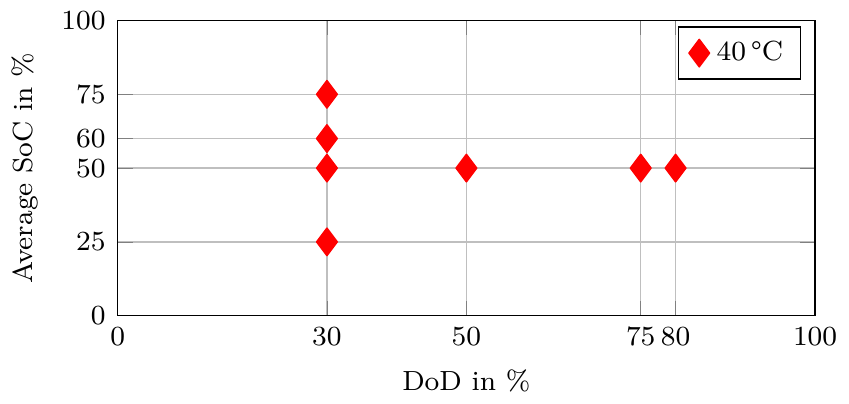}
		\caption{Cycle aging test conditions.}
		\label{fig:cycletestconditions}	
\end{figure}

In order to account for the aging of the EV battery cell we evaluated accelerated aging tests of Li-Tec HEA40 cells. These cells are identical in composition and construction to the Li-Tec HEA50 cells apart from the capacity and dimensions. Therefore, the results are transferable to the Li-Tec HEA50 cell of the Smart e.d. (3rd Generation).\\
The aging process of a battery cell leads to a reduction in capacity and an increase of its inner resistance. Calendar and cycle aging were considered separately. This approach and the parameterization process is described in detail by Schmalstieg in \cite{Schmalstieg.2014}. The aging behavior of the Li-Tec HEA40 cell was tested in accelerated aging tests in the laboratory at the institute for power electronics and electrical drives (ISEA) at RWTH Aachen. In order to separately measure the effects of calendar and cycle aging factors, two separate test procedures were carried out. The test conditions are shown in Fig. \ref{fig:caltestconditions} and Fig. \ref{fig:cycletestconditions}. Periodically, every 30 - 50 days, each cell underwent a checkup. During a checkup the capacity of the cell was measured in a 1 C discharge and the inner resistance was evaluated from the voltage response of the cell to a 1 C current charge pulse after $\SI{10}{\second}$. The overall aging of the cell is given by the superposition of calendar and cycle aging. \\

\subparagraph{Calendar Aging}
Calendar aging occurs at all times during storage and operation.  In the calendar aging test the battery cells were stored in climate chambers at a constant cell temperature and cell voltage. The test conditions for calendar aging are shown in Fig. \ref{fig:caltestconditions}. Three cells were tested for each test condition and their performance was averaged.  %
\\
\subparagraph{Cycle Aging}
In contrast to calendar aging, cycle aging occurs only when the cell is charged and discharged (cycled). During cycling the intercalation and de-intercalation of lithium ions leads to volume changes of the electrode material. This in turn can lead to crack-and-repair of the solid-electrolyte-interface (SEI) that consumes lithium (capacity loss) and increases its inner resistance. Also active material particles can loose contact to the electrode (capacity loss). \cite{Schmalstieg.2014}.\\
During the cycle aging test the battery cells were cycled (charged and discharged) with a current of 1 C ($\SI{50}{\ampere}$) and the cell temperature was kept constant at a temperature of $\SI{40}{\degreeCelsius}$ in a climate chamber. The test conditions are shown in Fig. \ref{fig:cycletestconditions}. One cell was tested for each condition. During the cycle aging tests also calendaric aging occurs which has to be accounted for during the fitting process.

\subsection{Charger and BMS Charge Control}
\label{sec:chargerbms}
\begin{figure}[H]
	\centering
	\includegraphics[width= \columnwidth]{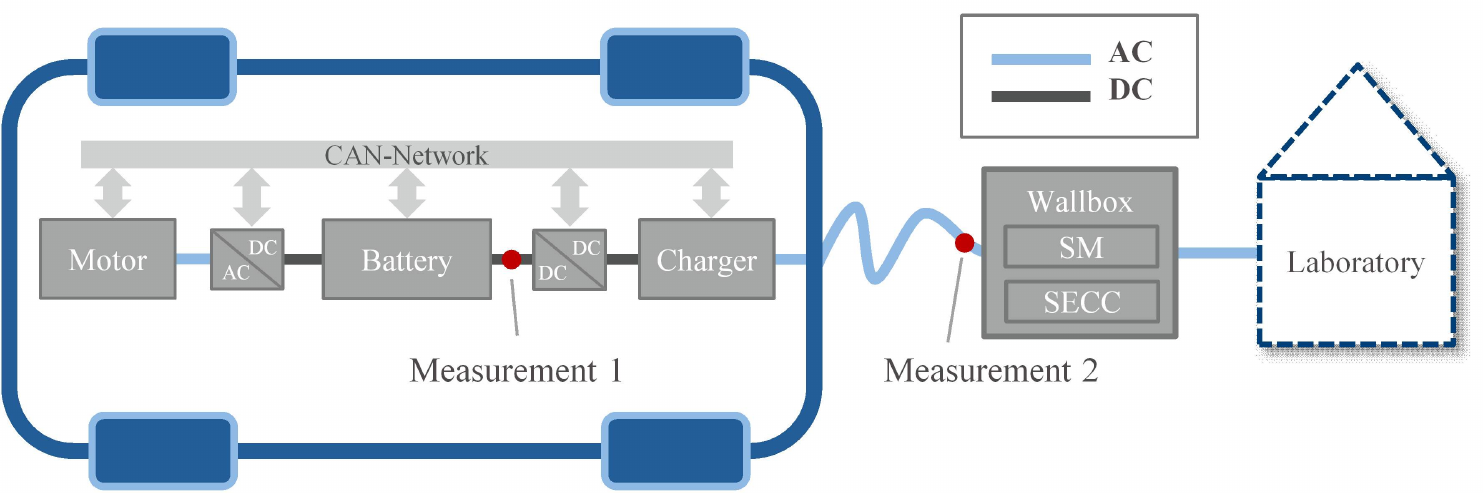}
	\caption{Laboratory setup for charger efficiency measurement}
	\label{fig:labsetupcharger}
\end{figure}

In Fig. \ref{fig:labsetupcharger} the laboratory setup for the parameterization of the charger and the controller for the charging process is shown. Two measurement points were used. At measurement point 1 the battery voltage and the battery current were measured. For this measurement the internal measurement devices of the vehicle were used which broadcast the values via the CAN-Bus. Via a CAN-Bus interface, the communication and therefore the measurement values were recorded. Measurement 2 was carried out on the grid side with a smart meter. 
The EV has an in-built 3-phase AC charger with a maximum charging power of \SI{22}{\KW}. The wallbox in the test setup has a Type 2 socket at which EVs can be charged via Mode 3 of IEC 61851-1. The maximum rated charging power is \SI{11}{\KW} (\SI{16}{\ampere}, 3-phase). The supply equipment communication controller (SECC) of the wallbox was controlled and monitored via a Modbus-TCP interface. Via this interface the maximum current can be set which the SECC transmits to the EV via pulse-width modulation (PWM) in accordance to IEC 61851-1. The SECC that was used, was only able to set integer values for the maximum charging current. Therefore, the charging current could only be increased/decreased in \SI{1}{\ampere} steps starting at a minimal current of \SI{6}{\ampere}.\\
As it was not possible to switch from 3-phase to 1-phase charging mode via the SECC for the Smart e.d., the measurements for 1-phase charging were carried out with the emergency charging cable of the Smart e.d. without the use of the wallbox. The emergency charging cable plugs into a Schuko (protective contact) socket and enables 1-phase charging via Mode 2. The charging current is set by an in-cable communication controller. The controller has two settings which set two different charging power settings: \SI{1.8}{\KW} and \SI{2.9}{\KW}.
The possible set-points for the charging process in the laboratory are summarized in Table \ref{tab:setpoints}.

\begin{table}[H]
	\centering
	\resizebox{0.5\textwidth}{!}{
		
		\begin{tabular}{cc}
			\toprule
			\textbf{Charging mode}	& \textbf{Possible setpoints from \SI{6}{\ampere} to \SI{16}{\ampere}}\\
			\midrule
			1-phase charging & \SI{1.8}{\KW}  and \SI{2.9}{\KW}\\
			3-phase charging & \SI{4.1}{\KW} - \SI{11}{\KW} in \SI{690}{\watt} steps \\
			\bottomrule
		\end{tabular}
		}
	\caption{Charging power setpoints for a grid voltage of \SI{230}{\volt} and the Smart e.d. of the wallbox (3-phase charging) and the emergency charging cable (1-phase charging).}
	\label{tab:setpoints}
\end{table}

%% file: sections/results.tex
\section{Results}
\label{sec:results}
In this section we show the results of the EV model parameterization.

\subsection{Electrical Model Parameters}
\label{sec:resultselectricalmodel}
\begin{figure*}[!htbp]
	\begin{subfigure}[t]{\columnwidth}
		\centering
		\includegraphics[]{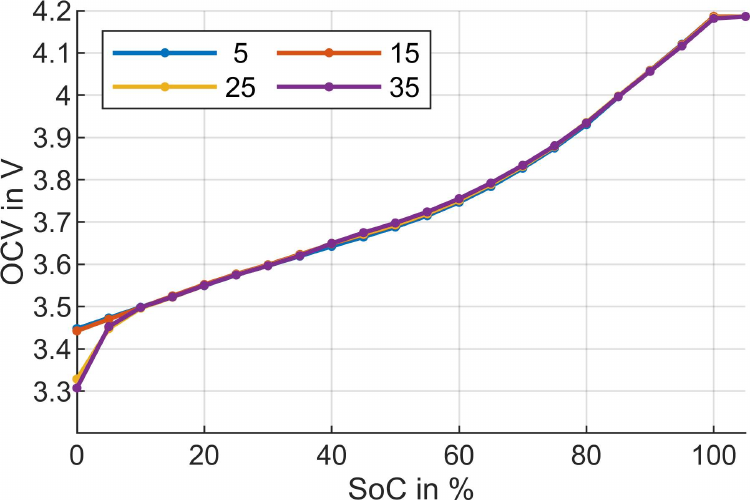}
		\caption{Open-circuit voltage (OCV).}
		\label{fig:ocvl}
		\vspace{6pt}
	\end{subfigure}
	\begin{subfigure}[t]{\columnwidth}
		\centering
		\includegraphics[]{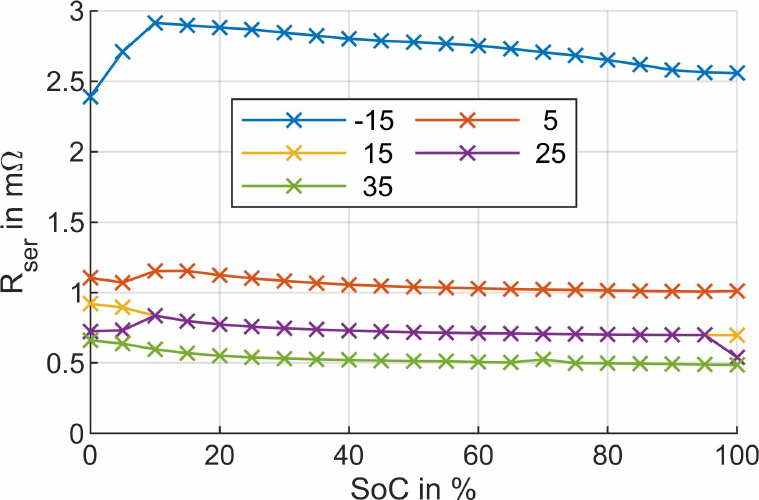}
		\caption{Resistance $R_{ser}$.}
		\label{fig:R_ser}
	\end{subfigure}
	\begin{subfigure}[t]{\textwidth}
		\centering
		\includegraphics[]{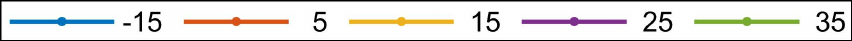}
		\vspace{6pt}
	\end{subfigure}
	\begin{subfigure}[t]{\columnwidth}
		\centering
		\includegraphics[]{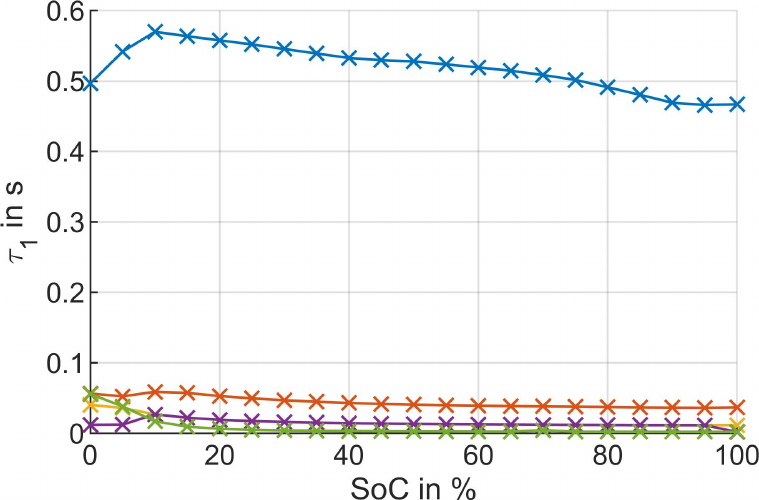}
		\caption{Time constant $\tau_{1} = R_{1}\cdot C_{1}$.}
		\label{fig:tau_1}
		\vspace{6pt}
	\end{subfigure}
	\begin{subfigure}[t]{\columnwidth}
		\centering
		\includegraphics[]{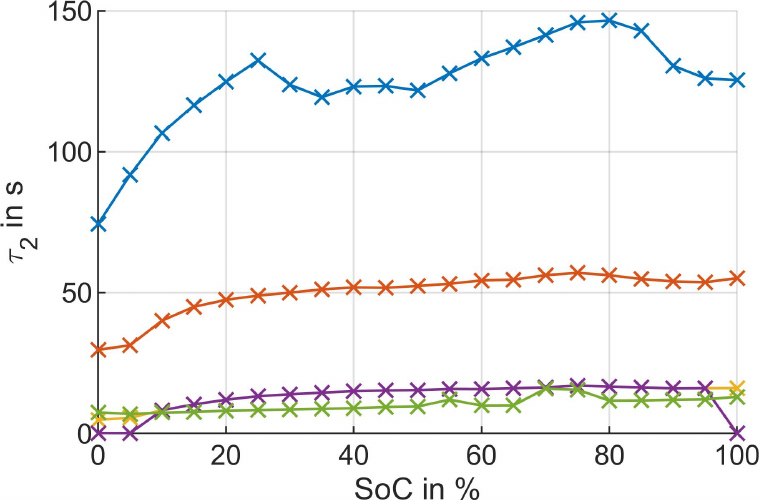}
		\caption{Time constant $\tau_{2} = R_{2}\cdot C_{2}$.}
		\label{fig:tau_2}
	\end{subfigure}
	\begin{subfigure}[t]{\columnwidth}
		\centering
		\includegraphics[]{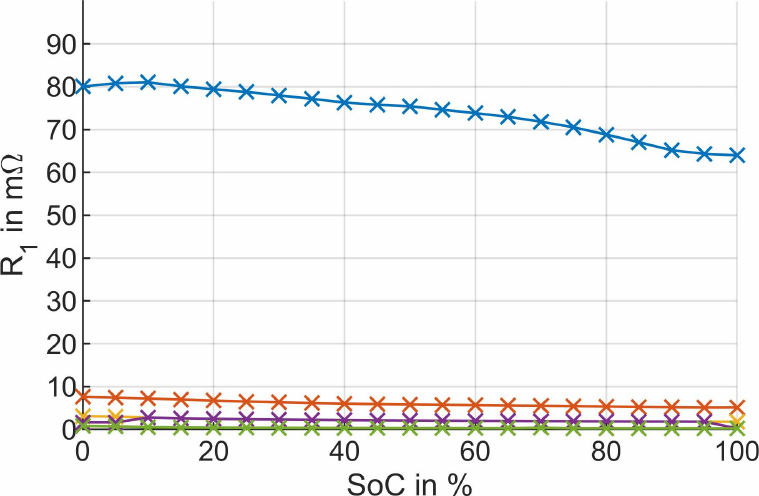}
		\caption{Resistance $R_{1}$.}
		\label{fig:R_1}
		\vspace{6pt}
	\end{subfigure}
	\begin{subfigure}[t]{\columnwidth}
		\centering
		\includegraphics[]{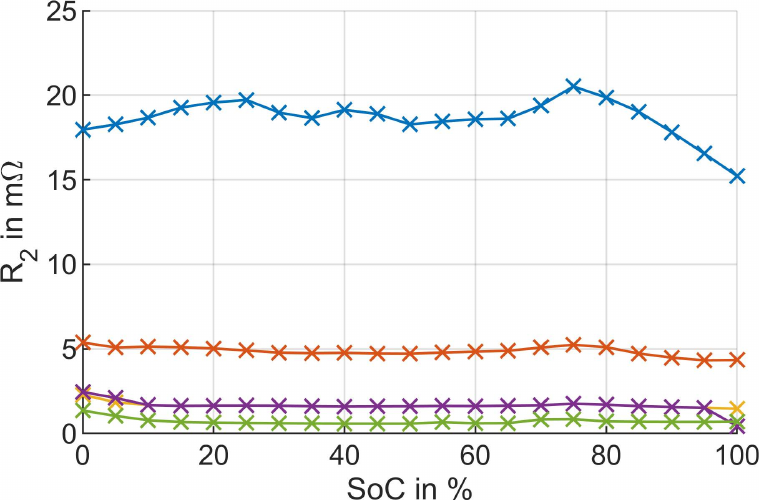}
		\caption{Resistance $R_{2}$.}
		\label{fig:R_2}
	\end{subfigure}
	\caption{Electrical parameters of  Li-Tec \SI{52}{\AH} \ce{LiNiMnCoO2} pouch cell}
	\label{fig:elparameters}
\end{figure*}

  Impedance spectra at different SOCs and temperatures of the cell were measured and the parameters for the ECM (see Fig. \ref{fig:eqd-cell}) were extracted from resulting Nyquist diagrams. Also, OCV measurements of the battery cell were conducted at different cell temperatures. The results are shown in Fig. \ref{fig:elparameters}. Resistance $R_{ser}$ shows little dependency of the SOC which indicates that no transient polarization process is occurring. Furthermore, $R_{ser}$ varies approximately half an order of magnitude with the cell temperature. The increased resistance at higher temperatures correlates with the underlying chemical process of an increased reaction rate. This behavior is also visible for $R_{1}$ and $R_2$ and proves the physical interpretability of these parameters. %
  Besides the resistances other important parameters to evaluate the parameterization of the ECM and the correspondence with physical processes are the time constants $\tau_{1}$ and $\tau_{2}$ of the RC elements. In our model the time constants $\tau_{1}$  and $\tau_{2}$  model the duration of the reaction and balancing processes, respectively. Thus the following has to apply: 
  \begin{equation}
  	\tau_1 =R_1 \cdot C_1 < R_2 \cdot C_2 = \tau_2
  	\label{eq:tauineq}
  \end{equation}
The results show the time constant $\tau_{1}$ to be in the range of $\SI{e-2}{\second}$ to $\SI{e-1}{\second}$ while $\tau_{2}$ in the $\SI{e1}{\second}$ to $\SI{e2}{\second}$ range. This is in accordance to the previously discussed physical interpretation and inequality \ref{eq:tauineq}.
For the simulation of charging and discharging processes (V2G) with the simulation mode these time constants fit the dynamic processes. Impedance components in the ECM with smaller time constants are not necessary. The ECM and parameterization is therefore appropriate for prosumer household simulations and V2G applications.

\subsection{Thermal Model Parameters}
\label{sec:resultsthermalmodel}
\begin{figure*}[!htbp]
	\begin{minipage}[b]{0.4\textwidth}
		\centering
		\subcaptionbox{Volume distribution of modules.\label{fig:volmod}}{\includegraphics[]{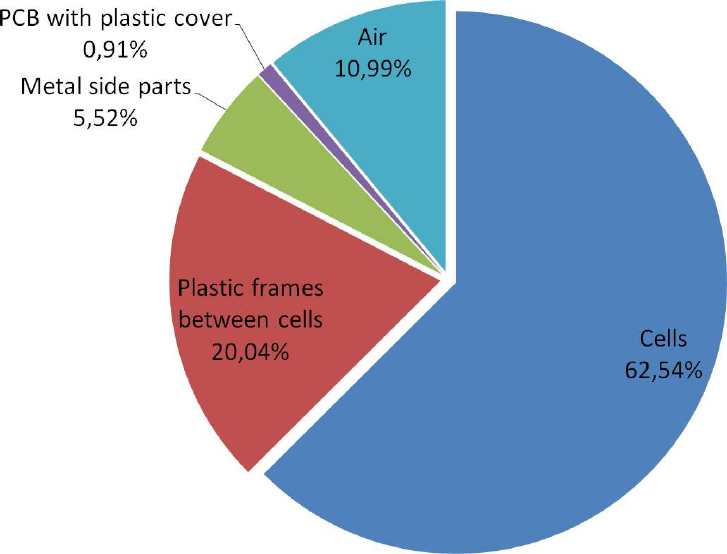}}
		\vspace{0.5cm}
	\end{minipage}
	\begin{minipage}{0.02\textwidth}
		\hfill
	\end{minipage}
	\begin{minipage}[b]{0.48\textwidth}
		\centering
		\subcaptionbox{Weight distribution of modules.\label{fig:weightmod}}{\includegraphics[]{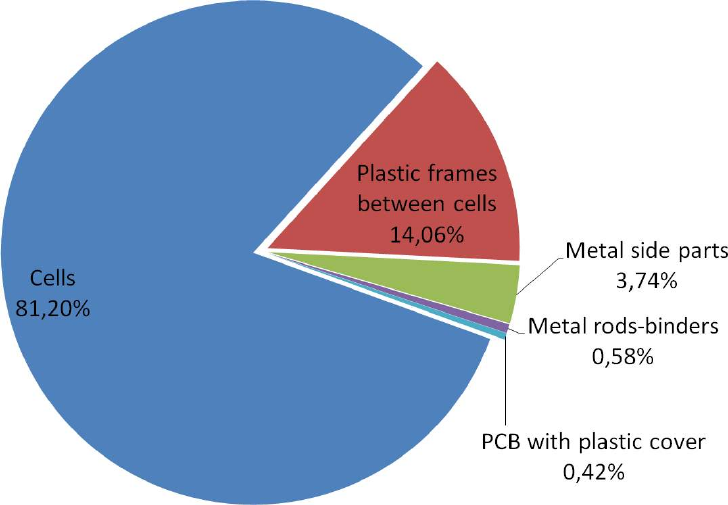}}
		\vspace{0.5cm}
	\end{minipage}
	\begin{minipage}[b]{0.48\textwidth}
		\centering
		\subcaptionbox{Volume distribution of pack.\label{fig:volpack}}{\includegraphics[]{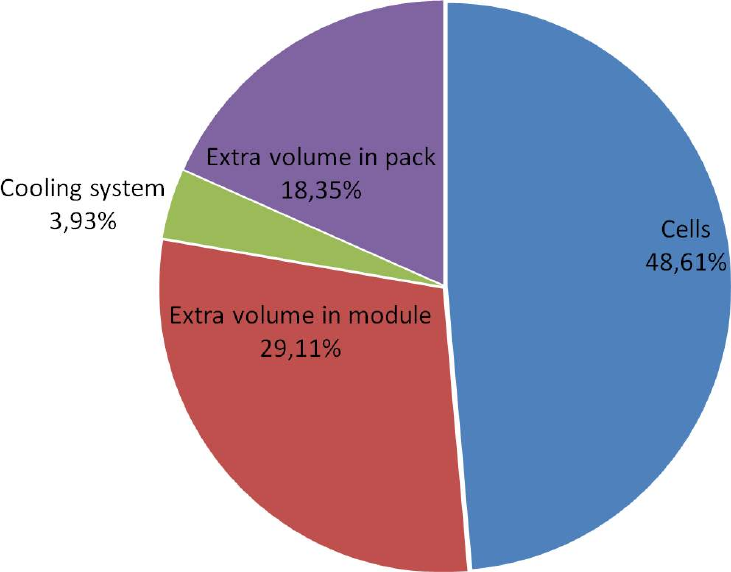}}
		\vspace{0.5cm}
	\end{minipage}
	\begin{minipage}{0.02\textwidth}
		\hfill
	\end{minipage}
	\begin{minipage}[b]{0.48\textwidth}
		\centering
		\subcaptionbox{Weight distribution of pack.\label{fig:volpack}}{\includegraphics[]{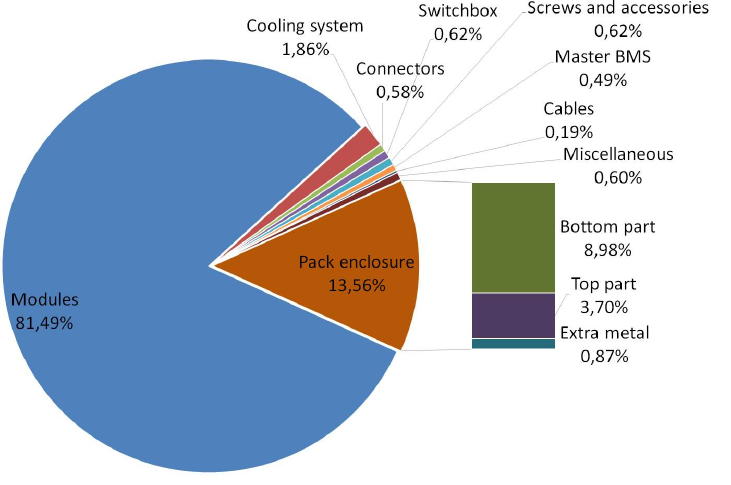}}
		\vspace{0.5cm}
	\end{minipage}
	\caption[Volume and weight distributions of the Smart e.d. battery pack]{Volume and weight distributions of the Smart e.d. battery pack} 
	\label{fig:packdistributions}
\end{figure*}

We disassembled a battery pack of the Smart e.d. and measured its dimensions (see table \ref{tab:packdimensions}). We also determined the weight and volume distributions of the pack and a single module (see Fig. \ref{fig:packdistributions})).
\begin{table}[H]
	\begin{tabular}{ccc}
		\toprule
		& Pack & Modules with slave BMS  \\ 
		Volume in $\SI[per-mode=symbol]{}{\liter}$ & 95.86 & 74.5  \\
		Mass in $\SI[per-mode=symbol]{}{\kilogram}$	& 179.6 & 146.5 \\
		Surface area in $\SI[per-mode=symbol]{}{\meter^2}$ & 1.69 & 1.36 \\ \bottomrule \hline
	\end{tabular}
	\caption{Traction battery pack and module dimensions.}
	\label{tab:packdimensions}
\end{table}
 We calculate the thermal parameters for the thermal equivalent circuit diagram (shown in Fig.\ref{fig:eqc:thermal}) using literature values for the specific heat capacity (shown in table \ref{tab:thermalspecs}) and the pack dimensions and weight distributions shown in table \ref{tab:packdimensions}. Different materials have different specific heat transfer coefficients to air and the battery pack is made up of different materials. Therefore, we fit the specific heat transfer coefficient for the open pack by performing simulations of the cooling phase of the pack without load after the 1C discharge measurement in the laboratory. The resulting specific heat transfer coefficient that provided the best agreement between simulation and data was $\alpha_{spec} = \SI{5}{\watt \per \meter^2 \per \kelvin}$. As the battery pack was not mounted in the car during the pack tests in the laboratory we obtain different heat transfer coefficients sets for the EV operation tests. The resulting parameters are shown in table \ref{tab:thermalresultspecs}.

\begin{table}[h]
	\centering
	\begin{tabular}{l l l}
		\toprule
		 & EV Operation & Lab Pack Test \\
		$C_{pack}$ in $\SI[per-mode=symbol]{}{\kilo \joule \per \kelvin}$ &  $\SI{17.12}{}$ & $\SI{17.12}{}$  \\
		$\alpha_x$ in $\SI[per-mode=symbol]{}{\watt \per \kelvin}$ & $\SI{0.726}{}$ & $\SI{0.472}{}$ \\ 
		$\alpha_y$ in $\SI[per-mode=symbol]{}{\watt \per \kelvin}$ & $\SI{3.470}{}$ &  $\SI{2.863}{}$ \\ 
		$\alpha_z$ in $\SI[per-mode=symbol]{}{\watt \per \kelvin}$ & $\SI{1.383}{}$ &  $\SI{0.899}{}$ \\ \midrule
	\end{tabular}
	\caption{Values for heat capacity of the pack  $C_{pack}$ and the convective heat transfer coefficients $\alpha$ for each spatial direction (x,y,z). Parameters of thermal battery pack model shown in Fig. \ref{fig:eqc:thermal}}
	\label{tab:thermalresultspecs}
\end{table}

\subsection{Aging Model Parameters}
\label{sec:resultsagingmodel}
We analysed the aging of the battery cells in the accelerated aging tests and fitted aging functions to the data.

\begin{figure}[H]
	\centering
	\includegraphics[width = \columnwidth]{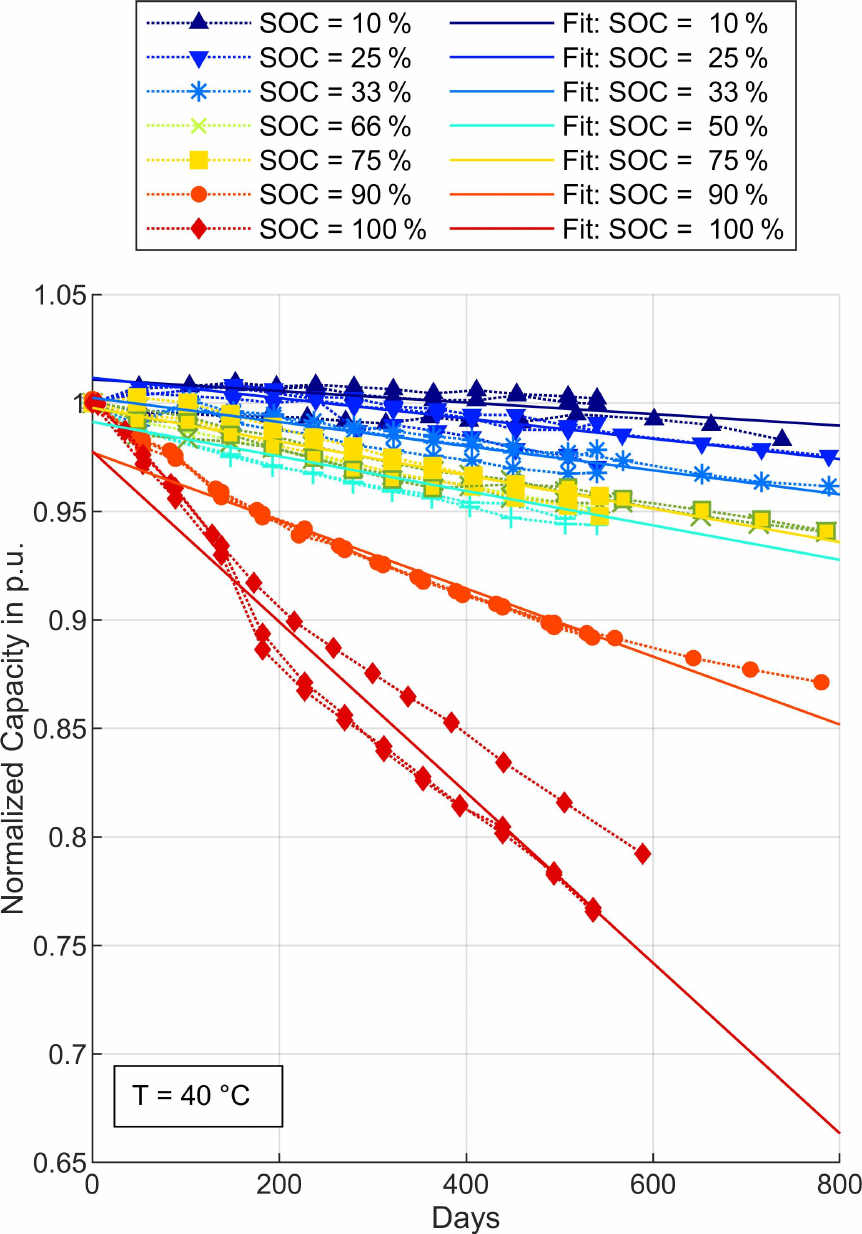}
	\caption{Capacity test and time fit results for cells stored at \SI{40}{\degreeCelsius} in the calendar aging tests.}
	\label{fig:agecapsoc}
\end{figure}

\begin{figure}[H]
	\centering
	\includegraphics[width = \columnwidth]{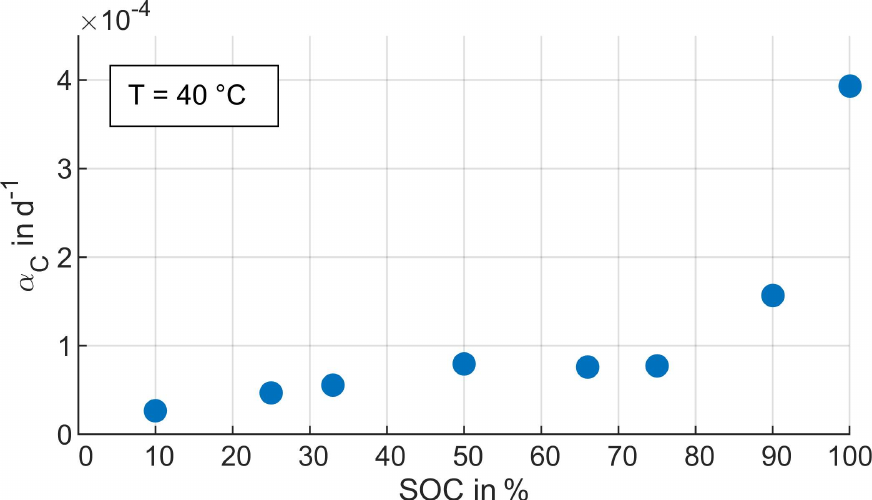}
	\caption{Linear coefficient $\alpha_{C}$ of the time fit results for cells stored at \SI{40}{\degreeCelsius} in the calendar aging tests.}
	\label{fig:agecapsocalpha}
\end{figure}

\begin{figure}[H]
	\centering
	\includegraphics[width = \columnwidth]{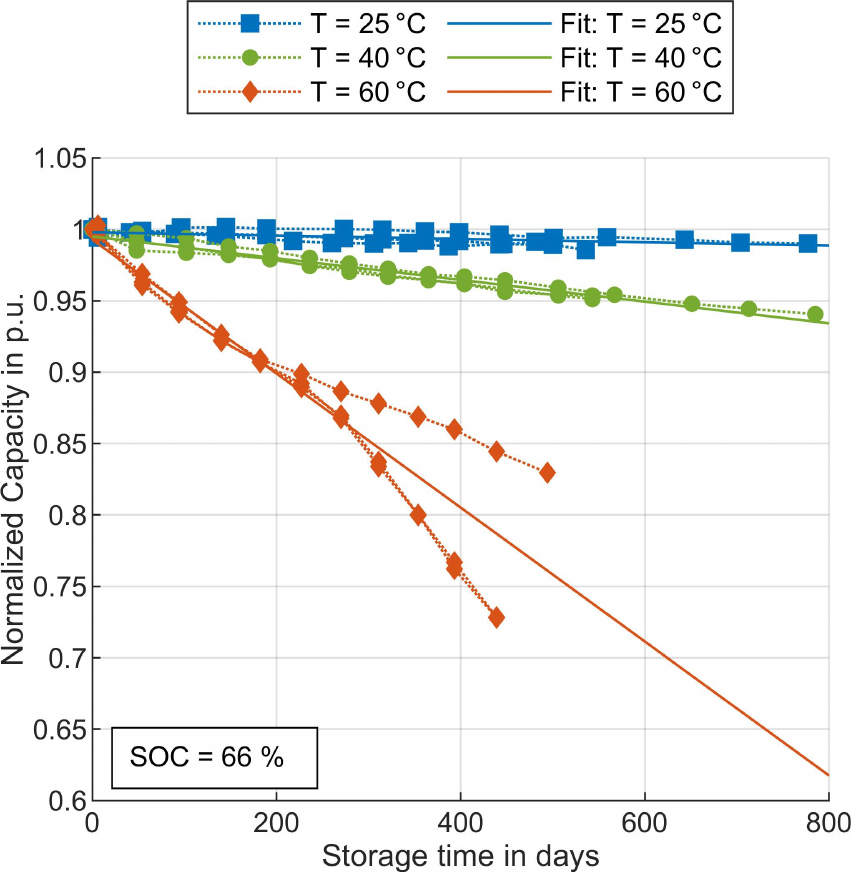}
	\caption{Capacity test and time fit results for cells stored at \SI{66}{\%} SOC in the calendar aging tests.}
	\label{fig:agecaptemp}
\end{figure}

\begin{figure}[H]
	\centering
	\includegraphics[width = \columnwidth]{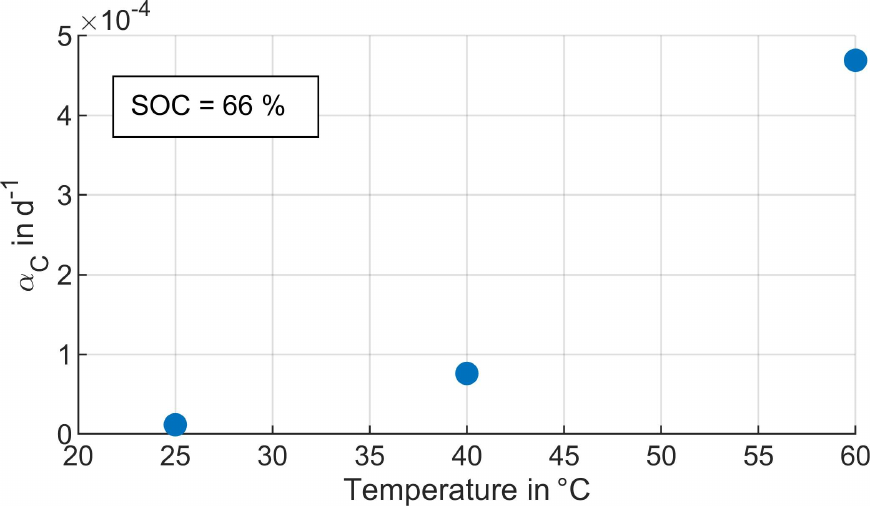}
	\caption{Linear coefficient $\alpha_{C}$ of the time fit results for cells stored at \SI{66}{\%} SOC in the calendar aging tests.}
	\label{fig:agecaptempalpha}
\end{figure}

\begin{figure}[H]
	\centering
	\includegraphics[width = \columnwidth]{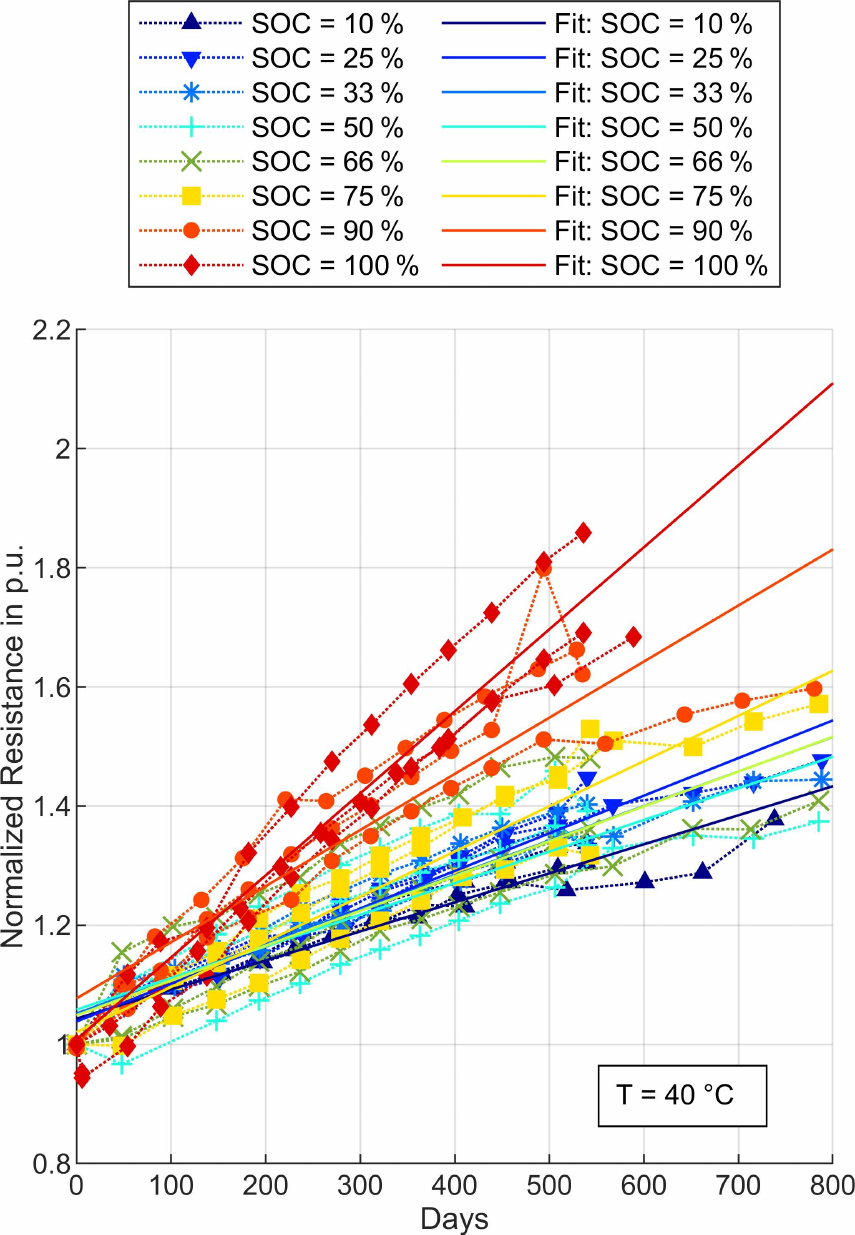}
	\caption{Inner resistance measurement and time fit results for cells stored at \SI{40}{\degreeCelsius} in the calendar aging tests.}
	\label{fig:ageressoc}
\end{figure}

\begin{figure}[H]
	\centering
	\includegraphics[width = \columnwidth]{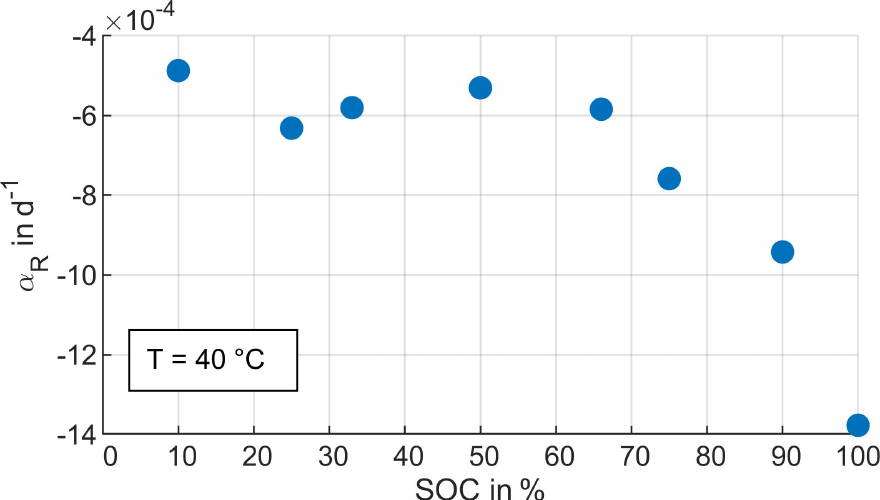}
	\caption{Linear coefficient $\alpha_{R}$ of the time fit results for cells stored at \SI{40}{\degreeCelsius} in the calendar aging tests.}
	\label{fig:ageressocalpha}
\end{figure}

\begin{figure}[H]
	\centering
	\includegraphics[width = \columnwidth]{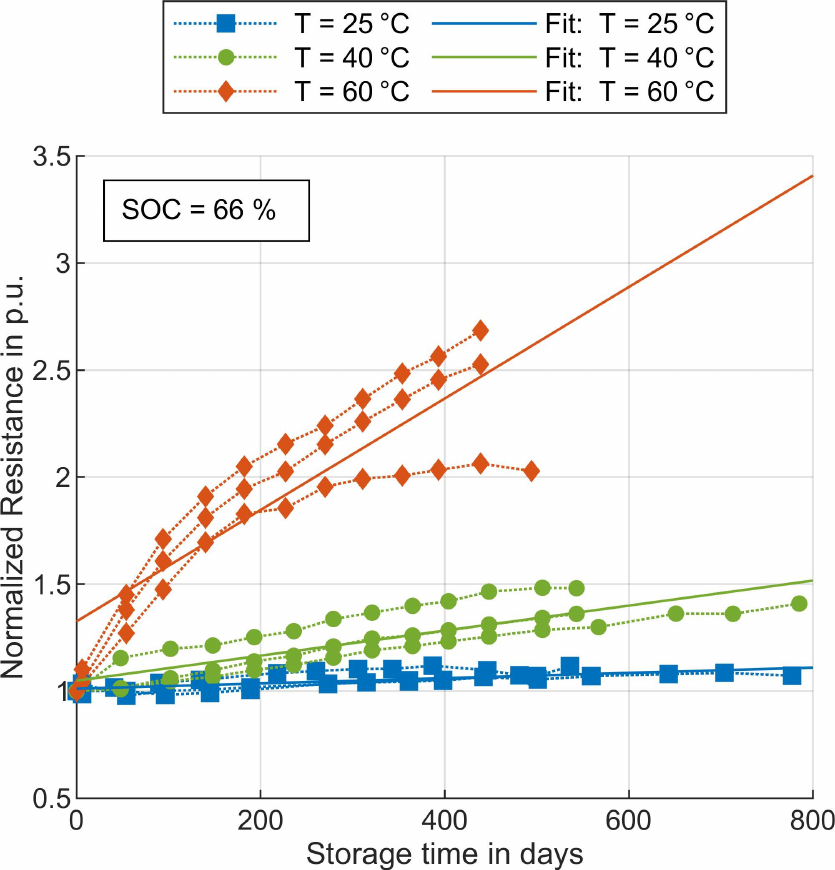}
	\caption{Inner resistance measurement and time fit results for cells stored at \SI{66}{\%} SOC  in the calendar aging tests.}
	\label{fig:agerestemp}
\end{figure}

\begin{figure}[H]
	\centering
	\includegraphics[width = \columnwidth]{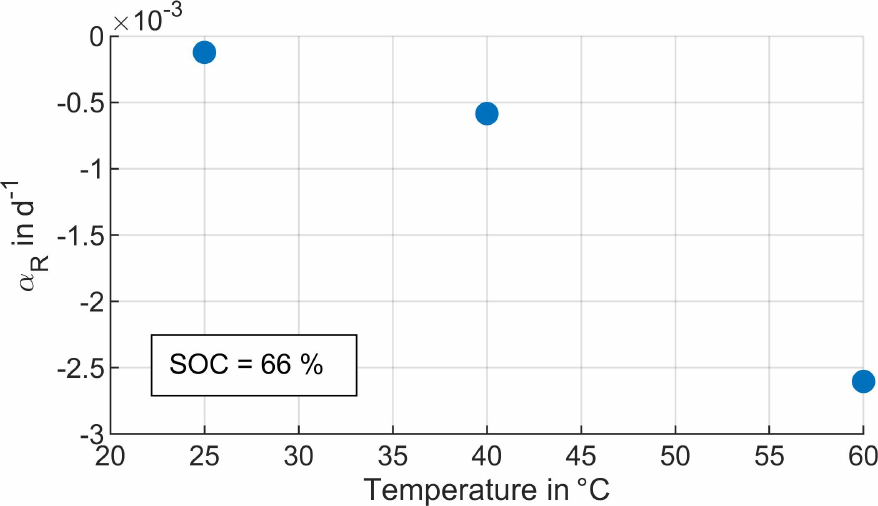}
	\caption{Linear coefficient $\alpha_{R}$ of the time fit results for cells stored at \SI{66}{\%} SOC  in the calendar aging tests.}
	\label{fig:agerestempalpha}
\end{figure}

\subparagraph{Calendar Aging}
The experimental data of the calendar aging test together with the time fit is shown in Fig. \ref{fig:agecapsoc} and Fig. \ref{fig:agecaptemp} for the normalized capacity and in Fig. \ref{fig:ageressoc} and Fig. \ref{fig:agerestemp} for the normalized resistance of the cell. As the cells age, their capacity decreases and their resistance increases. For each set point we tested three cells. The figures show that the aging characteristic of three cells at the same set point can differ to a large extent. In a previous study this was linked to variances of material properties and process parameters in the production process. \cite{BAUMHOFER2014332} 
In order to fit the time dependence for calendar aging we averaged the experimental results for each set point and fit eq. \ref{eq:agingcap} and eq. \ref{eq:agingres} for capacity and resistance respectively:
\begin{align}
	C_{norm} &= 1-\alpha_{C} \cdot t \label{eq:agingcap}\\
	R_{norm} &= 1+\alpha_{R} \cdot t, \label{eq:agingres}
\end{align}
with capacity and resistance given as normalized parameters:
\begin{align}
	C_{norm} &= \frac{C}{C_{0}} \\
	R_{norm} &= \frac{R}{R_{0}},	
\end{align}
where $C_0$ is the initial capacity and $R_0$ is the initial inner resistance.\\
Other exponents for the time dependency can be found in literature, such as 0.75, which was found to describe the time dependence for calendar aging in \cite{Schmalstieg.2014}. However, the linear approach yielded the best fit overall and was therefore chosen for this cell. 
The coefficients $\alpha_{C}$ are shown in Fig. \ref{fig:agecapsocalpha} and Fig. \ref{fig:agecaptempalpha}. The coefficients $\alpha_{R}$ are shown in Fig. \ref{fig:ageressocalpha} and Fig. \ref{fig:agerestempalpha}.
The results for calendar aging at \SI{40}{\degreeCelsius} show accelerated loss of capacity of cells stored at higher SOCs with a plateau  between 50 and 75 \%. %
The results for the cell resistance show a plateau for cells stored between 25 \% and 75 \% SOC.
The capacity of cells stored at 100 \% SOC decreases at a 112 \% higher rate than the capacity of cells stored at 90 \% which in turn decreases at a 150 \% higher rate than the capacity of cells stored at 75 \% SOC. The highest rate of increase of the inner resistance exhibited cells stored at 90 \% and 100 \% SOC (see Fig. \ref{fig:ageressoc}). However, the cell resistance never reached the end-of-life criterion of a 100 \% increase of the inner resistance.  
Figures \ref{fig:agecaptemp} and \ref{fig:agerestemp} show the calendar aging test results for cells stored at  66 \% SOC and different temperatures. The aging rate of the cell depends strongly on the cell temperature. The rate of capacity decrease at an SOC of  66 \%  is 470 \% higher at \SI{60}{\degreeCelsius} than at \SI{40}{\degreeCelsius} which in turn leads to a 480 \% higher rate than for cells stored at \SI{25}{\degreeCelsius}. The results follow the same trend as for the inner resistance. The cells stored at \SI{60}{\degreeCelsius} and 66 \% SOC reached the end-of-life criterion of the inner resistance first after 200 to 300 days. In conclusion, avoidance of high cell temperatures ($<  \SI{60}{\degreeCelsius}$) and high storage SOCs ( $> 75 \%$) is highly beneficial to reduce calendar aging of this cell.





\subparagraph{Cycle Aging}

\begin{figure}[H]
	\centering
	\includegraphics[width = \columnwidth]{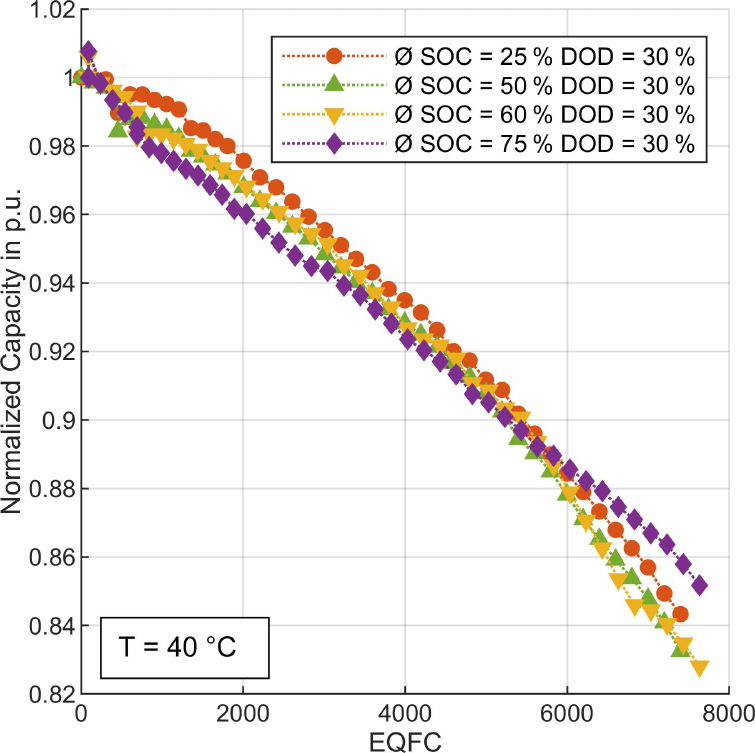}
	\caption{Capacity test results for cells cycled with \SI{30}{\%} in the cyclic aging tests ($T=\SI{40}{\degreeCelsius}$).}
	\label{fig:agecyccapsoc}
\end{figure}

\begin{figure}[H]
	\centering
	\includegraphics[width = \columnwidth]{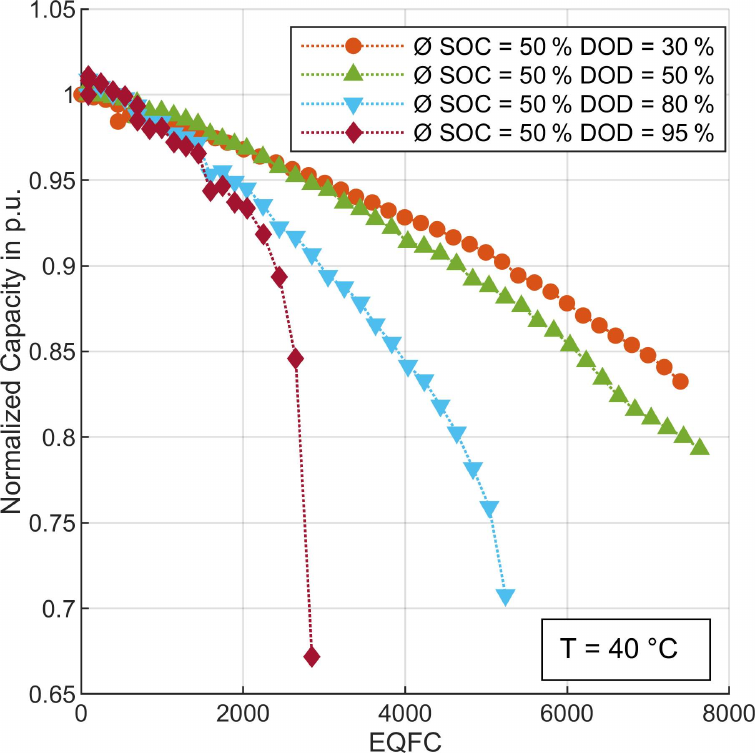}
	\caption{Inner resistance measurement results for cells cycled around \SI{50}{\%} in the cyclic aging tests ($T=\SI{40}{\degreeCelsius}$).}
	\label{fig:agecyccapdod}
\end{figure}

\begin{figure}[H]
	\centering
	\includegraphics[width = \columnwidth]{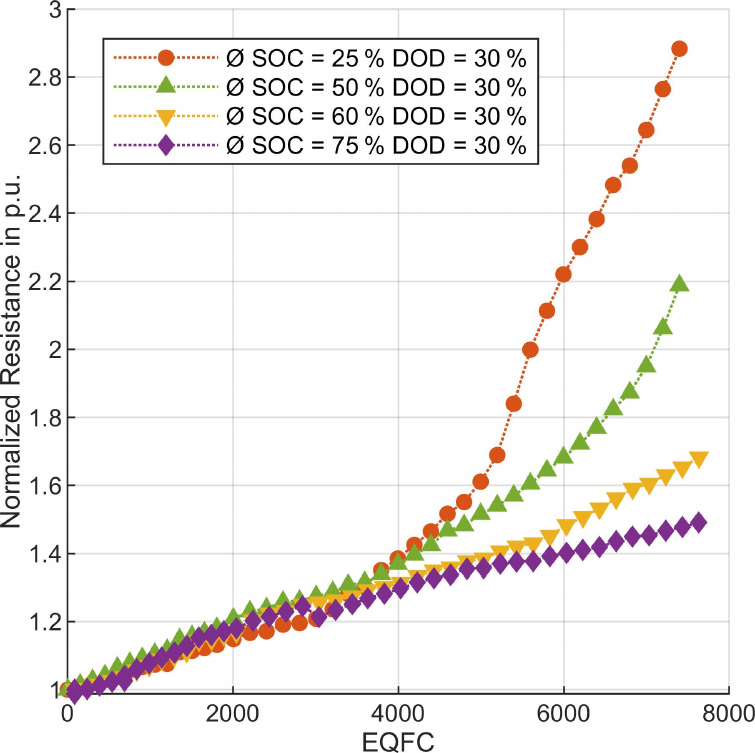}
	\caption{Inner resistance measurement results for cells cycled with \SI{30}{\%} in the cyclic aging tests ($T=\SI{40}{\degreeCelsius}$).}
	\label{fig:agecycressoc}
\end{figure}

\begin{figure}[H]
	\centering
	\includegraphics[width = \columnwidth]{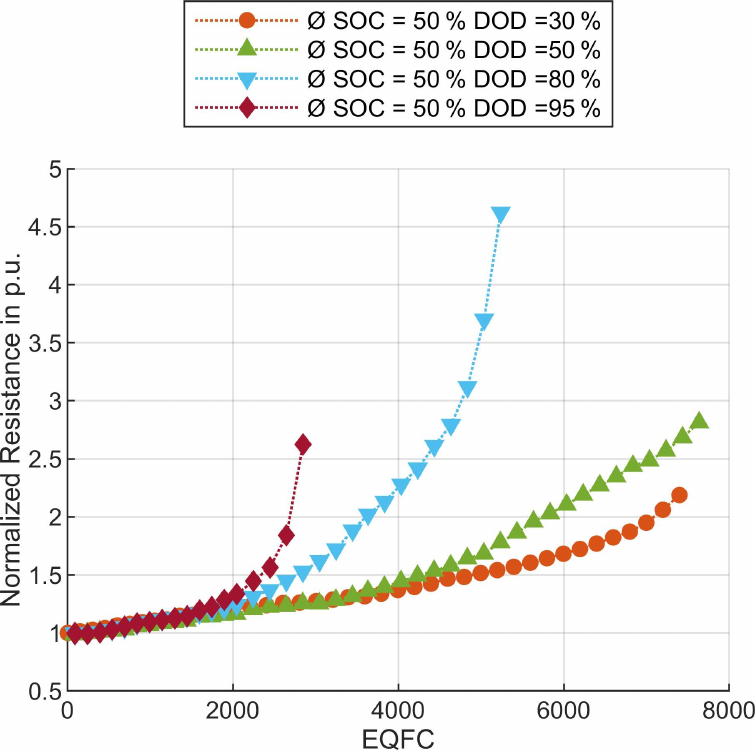}
	\caption{Capacity test results for cells cycled around \SI{50}{\%} in the cyclic aging tests ($T=\SI{40}{\degreeCelsius}$).}
	\label{fig:agecycresdod}
\end{figure}

The results of the cycle life tests are shown in Figures \ref{fig:agecyccapsoc} and \ref{fig:agecyccapdod} for the capacity and Figures \ref{fig:agecycressoc} and \ref{fig:agecycresdod} for the inner resistance. All tests were carried out at  \SI{40}{\degreeCelsius}. The cells in the cycle life tests age due to calendar aging and cycle aging.%
The results of tests carried out with different DOD show that the lifetime of the cells decrease with increasing DOD. All cells reached one of the two end-of-life criteria (80 \% of initial capacity and/or 200 \% of initial inner resistance) with the exception of the cells cycled with a DOD of 30 \% around a mean SOC of 60 \% and 75 \%. The first cell to reach EOL was the cell that was cycled with a DOD of 95 \% around a mean SOC of 50 \%. The EOL was reached in the EQFC range of 2649 to 2849. %
The second cell to reach EOL is the cell that was cycled with 80 \% DOD around an SOC of 50 \%. The EOL is reached after 3634 EQFC. %


\subsection{Validation of traction battery model}
\begin{figure}[H]
	\centering
	\includegraphics[width = \columnwidth]{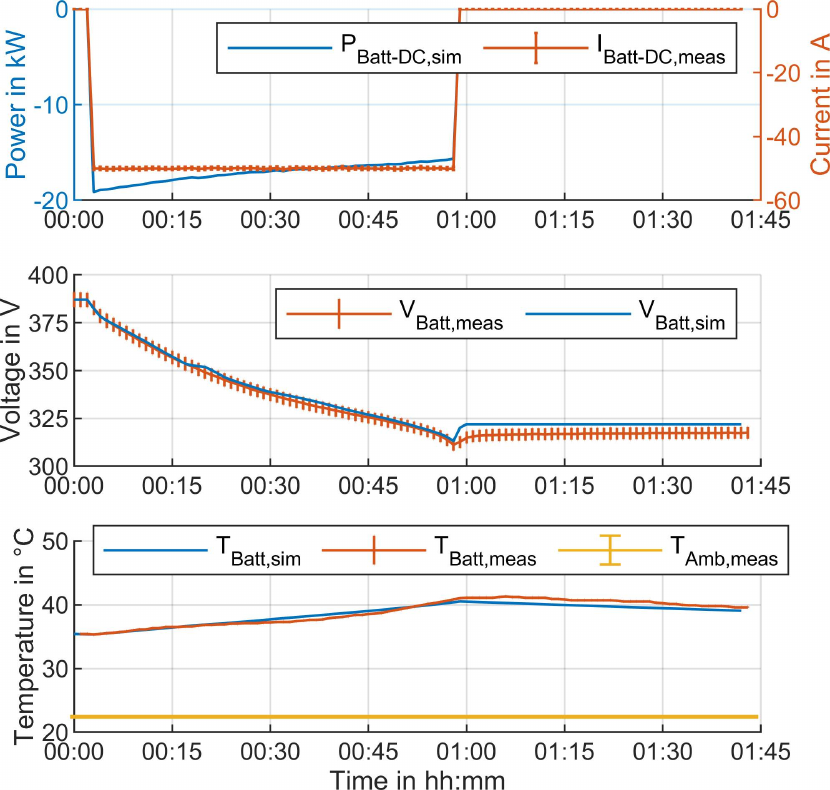}
	\caption{Laboratory pack test measurements and simulation: Full 1C Discharge.}
	\label{fig:packtestdischarge}
\end{figure}

\begin{figure}[H]
	\centering
	\includegraphics[width = \columnwidth]{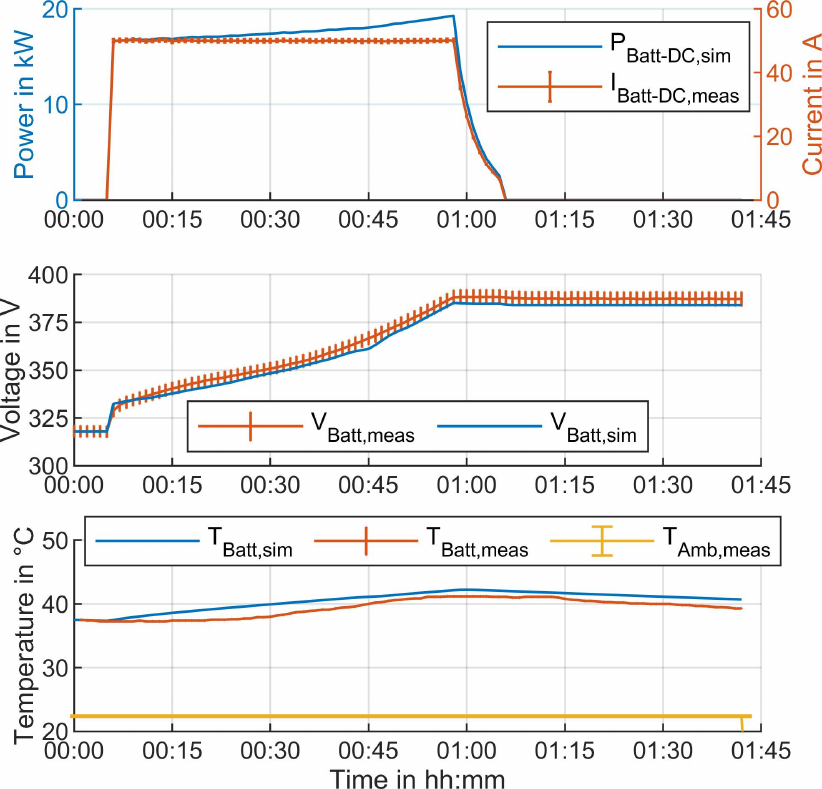}
	\caption{Laboratory pack test measurements and simulation: Full 1C Charge.}
	\label{fig:packtestcharge}
\end{figure}

\begin{figure}[H]
	\centering
	\includegraphics[width = \columnwidth]{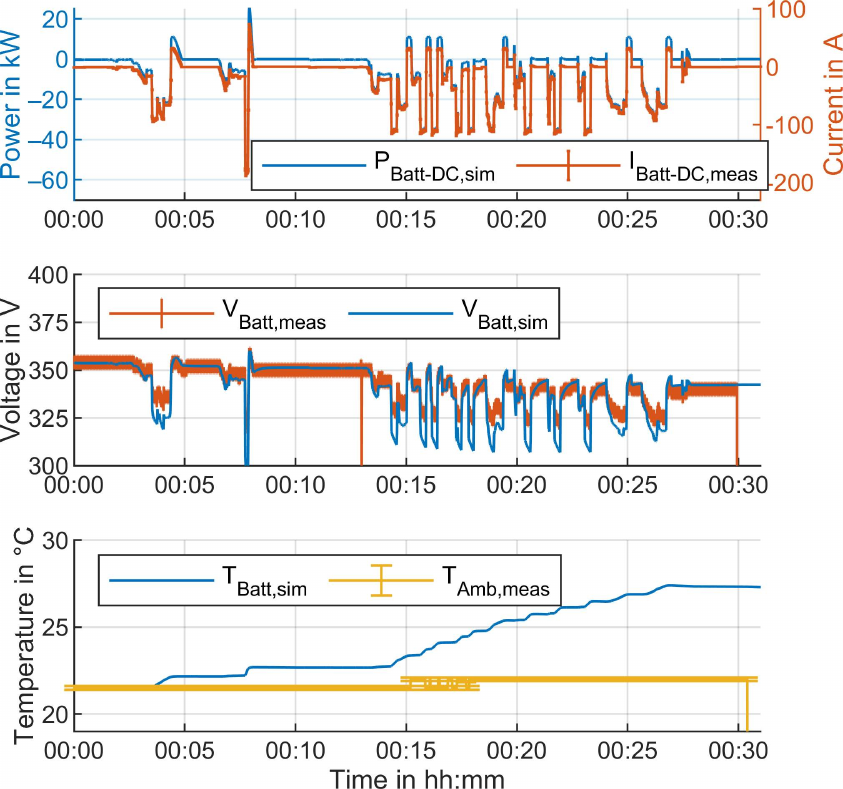}
	\caption{Driving profile measurement and simulation with \SI{1}{\second} resolution.}
	\label{fig:drivingprofile}
\end{figure}

\begin{figure}[H]
	\centering
	\includegraphics[width = \columnwidth]{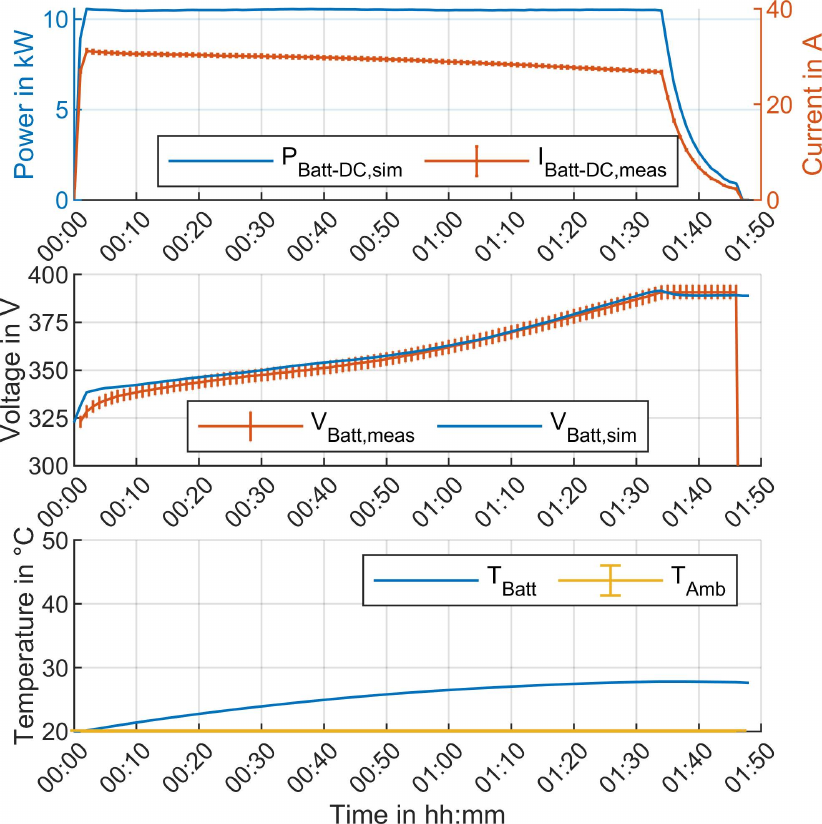}
	\caption{Measurement and simulation of \SI{11}{\kilo \watt} charge.}
	\label{fig:chargeprofile}
\end{figure}

In the research project GoELK several Smart e.d. were operated in commercial fleets. After approximately two years of operation in a fleet for geriatric care, the battery packs of two Smarts were tested and disassembled in the laboratory at ISEA (RWTH Aachen). The battery pack was removed from the EV, the lid was taken off the battery pack and the pack was connected to a Digatron Pack Test unit. During the operation of the pack using the Digatron Pack Test unit, the BMS and the liquid cooling system of the Smart were not activated. The pack voltage, each cell voltage, ambient temperature and cell temperatures were recorded. In Fig. \ref{fig:packtestdischarge} and Fig. \ref{fig:packtestcharge}, the measurement of a full discharge and charge test respectively are shown and compared to the model simulation results. As the lid had been removed and the liquid cooling system was not activated, the thermal model was parameterized with the values for \textit{Lab Pack Test} in table \ref{tab:thermalresultspecs}. In Fig. \ref{fig:drivingprofile} and Fig. \ref{fig:chargeprofile}, the measurement and simulation of a driving profile and a charge process with $\SI{11}{\kilo \watt}$ is shown. For these simulations the thermal model was parameterized with the values for \textit{EV Operation} in table \ref{tab:thermalresultspecs} for a normally operated EV (Smart e.d.).\\ In table \ref{tab:validationresults} the profile characteristic and the deviation between measurement and simulation are shown for the four profiles. Overall, the comparison between measurement and simulation shows good agreement. It should be noted that in the model the possible deviation in voltage, state of charge and state of health between the 93 cells is neglected. Especially in the case of the laboratory pack test measurements this could have a non negligible effect as the BMS, and therefore also cell balancing, is not operational. The RMSE per cell (93s1p configuration) between measurement and simulation is between $\SI{18.49}{\milli \volt}$ and $\SI{31.26}{\milli \volt}$ for the profiles with a low dynamic (charging and constant current profiles). The RMSE per cell for the driving profile was higher with $\SI{67.17}{\milli \volt}$, which is still sufficiently accurate. Furthermore, this was expected as the ECM of the battery cell was parameterized to show higher accuracy for long profiles with low dynamics as the EV spends longer times charging than driving in the prosumer simulations. %
In case of the pack temperature, the absolute error and RMSE are larger than for the cell voltage which was to be expected due to the use of a lumped reduced order model. The errors in the lab test however, showed acceptable accuracy. Furthermore, a privately used vehicle in Germany is parked for 97 \% of the time. \cite{infasInstitutfurangewandteSozialwissenschaftGmbH.2017} %
 The impact of inaccuracies of the thermal model during operation are therefore limited.\\ 
In summary, the simulation model yields accurate results for the operation of an EV in prosumer households.
\begin{table*}[h]
	\centering
	\begin{tabular}{l| l l| l l}
		\toprule
		& \multicolumn{2}{c}{Laboratory Pack Test} & \multicolumn{2}{c}{EV Operation} \\
		Profile & 1 C Discharge &1 C Charge  & Driving  & Charge with $\SI{11}{\kilo \watt}$ \\
		Figure & Fig. \ref{fig:packtestdischarge} & Fig. \ref{fig:packtestcharge} & Fig. \ref{fig:drivingprofile} & Fig. \ref{fig:chargeprofile} \\
		Charge in Ah  & $-46.46 \pm 0.47 $ & $46.54 \pm 0.47 $ & $-10.34 \pm 0.10 $ & $47.16 \pm 0.47$ \\
		Energy in kWh  & $-15.87 \pm 0.32 $ & $16.60 \pm 0.33 $ & $-3.41 \pm 0.07 $ & $16.9 \pm 0.34$ \\
		Duration in min & 55.95 & 60.78 & 29.88 & 105 \\ \hline
		\multicolumn{5}{l}{Cell voltage deviation:} \\
		RMSE in mV & 18.49 & 31.26 & 67.17 & 30.22 \\
		Max. abs. error in mV & 80.92 & 59.45 & 278.87 & 43.59 \\ \hline
		\multicolumn{5}{l}{Pack temperature deviation:} \\
		RMSE in K & 0.36 & 1.29 & n.a. & n.a. \\
		Max. abs. error in K & 1.05 & 1.99 & n.a. & n.a. \\ \bottomrule \hline
	\end{tabular}
	\caption{Summary of validation profile parameters and simulation deviations.}
	\label{tab:validationresults}
\end{table*}%

\subsection{Charger}
\label{sec:chargerresults}
In this section, we present the results of the charger efficiency and charge control parameterization.
\subsubsection{Charger efficiency}
\begin{figure}[H]
	\begin{subfigure}[t]{\columnwidth}
		\centering
		\includegraphics[width = \columnwidth]{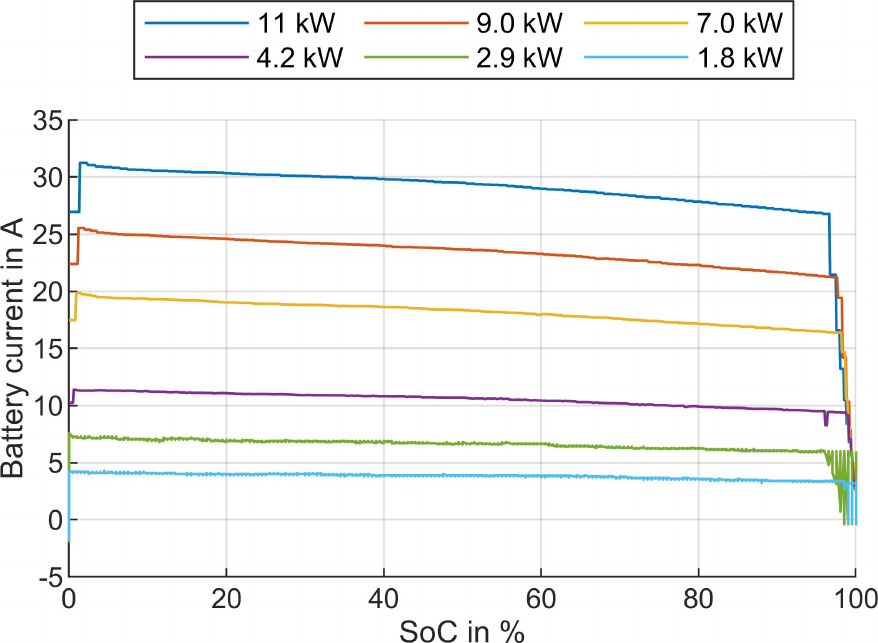}
		\caption{}
		\label{fig:chargecurve}
	\end{subfigure}
	\begin{subfigure}[t]{\columnwidth}
		\centering
		\includegraphics[width = \columnwidth]{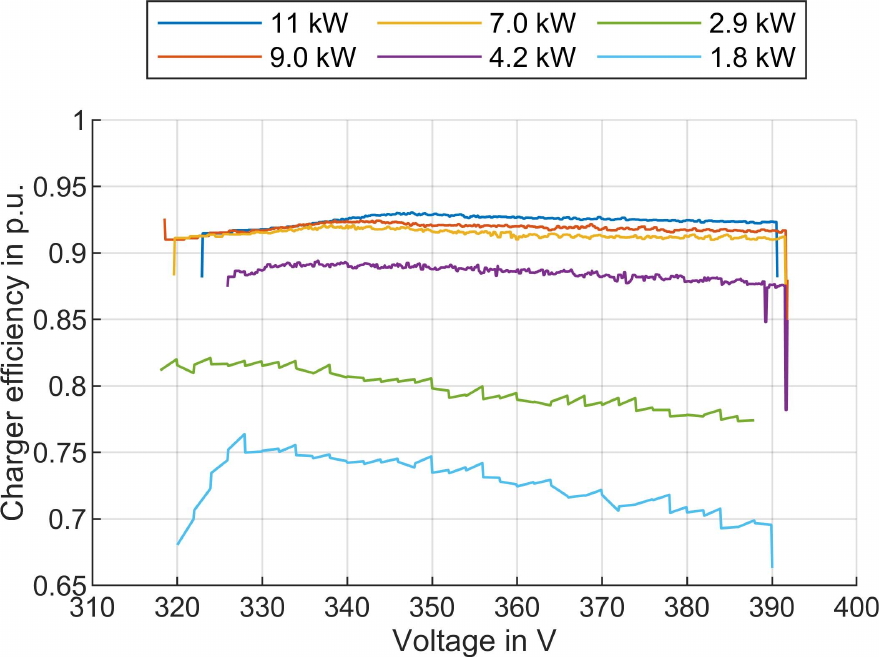}
		\caption{}
		\label{fig:efficiency}
	\end{subfigure}
	\caption{(a) Charging current (DC) versus SOC  using different charging power settings. (b) Charging efficiency versus battery voltage using different charging power settings}
	\label{fig:chargecurveefficiency}
\end{figure}

In Fig. \ref{fig:chargecurve}, the DC current of the traction battery at measurement point 1 (see Fig. \ref{fig:labsetupcharger}) is shown versus the SOC during charging processes. The power values denote the maximum power that was reached during the charging process. The curves for \SI{1.8}{\KW} and \SI{2.9}{\KW} were carried out with 1-phase whereas the other measurements were carried out with 3-phases. As the EV charges with constant power until the voltage limit is reached, the battery current decreases with increasing SOC. The sharp drop of the battery current at high SOC values is due to the constant voltage (CV) phase upon reaching the voltage limit. In Fig. \ref{fig:efficiency}, the efficiency of the charging process between AC power output of wallbox and DC power of battery (measurement points 2 and 1 respectively in Fig. \ref{fig:labsetupcharger} ) is shown versus the battery voltage for the same charging processes. The battery voltage is in the range of \SI{318.1}{\volt} - \SI{391.8}{\volt} for these measurements. %
The efficiency of the charger increases with the charging power. The efficiency for one full charge ranges from 73 \% for 1-phase charging with a setting of \SI{1.8}{\KW} to 92 \% for 3-phase charging with a setting of \SI{11}{\KW}. The lower values for the efficiency at low and high values of the battery voltage are due to low charging powers during the ramp up at the beginning of the charging process and the ramp down during the CV phase respectively.
\subsubsection{Charge control}
\begin{figure}[H]
	\begin{subfigure}[t]{\columnwidth}
		\centering
		\includegraphics[width = \columnwidth]{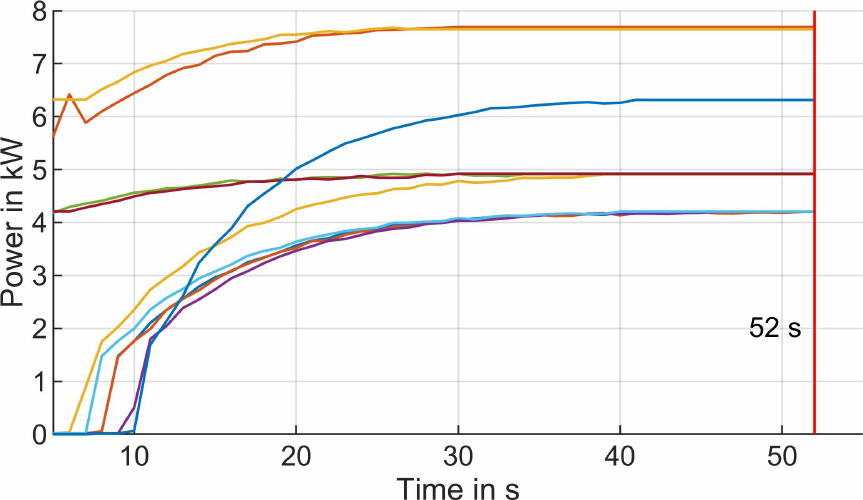}
		\caption{Charging power}\label{fig:delaya}
	\end{subfigure}
	\begin{subfigure}[t]{\columnwidth}
		\centering
		\includegraphics[width = \columnwidth]{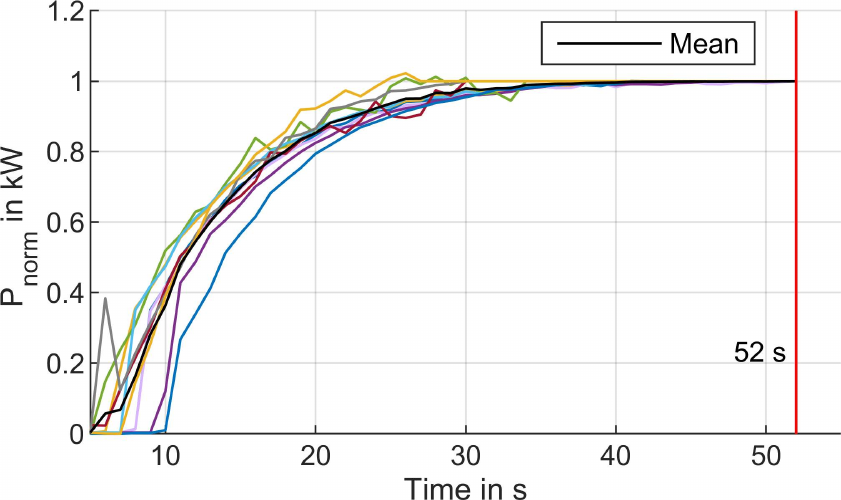}
		\caption{Normalized charging power}\label{fig:delayb}	
	\end{subfigure}
	\caption{Charging power (AC, measurement point 2) after a new maximum current setting is communicated to the EV via the CP at time t = \SI{0}{\second}. After \SI{52}{\second} the new charging power setting is reached by the EV.}
	\label{fig:delay}
\end{figure}

When the setting of the maximum charging current is changed via the CP, the new charging current is not reached by the EV immediately, but with a delay. This behaviour is shown in Fig. \ref{fig:delay} for the Smart. In Fig. \ref{fig:delaya}, the charging power on the AC side of the wallbox is shown after a new maximum current setting has been communicated to the EV via the CP at time $t=\SI{0}{\second}$.  The new power set-point $P_{EV,EMS}$ is not reached immediately but it takes some time for the power to ramp up. In all measurements the power set-point was reached after at most \SI{52}{\second}. The ramp up follows a similar curve for all measurements. Only in the case that the initial power set-point of the EV was \SI{0}{\watt}, we observed an initial reaction delay in the order of seconds.  We illustrate the similar behaviour of the EV upon a new power set-point with the normalized power $P_{norm}$ n Fig. \ref{fig:delayb}. We calculated it as follows:
\begin{align}
	P_{norm} = \frac{\left(P(t)-P(t=0)\right)}{P_{diff}} \\
	P_{diff} = P_{EV,EMS}-P(t=0)
\end{align}
In order to model the delay in the simulation we parameterize the average ramp up behavior $P_{ramp_{+}}$ using the average of all measurements $P_{norm,mean}(t)$ shown in Fig. \ref{fig:delayb}. We model the ramp up for a requested power increase according to Eq. \ref{eq:delayup} , where $t$ denotes the time after the the new power set point has been set.

\begin{equation}
	P_{ramp_{+}}(t) = 
	\begin{cases}
		P(t=0) + P_{diff} \cdot P_{norm,mean}(t)  & 0 \leq t \leq \SI{52}{\second}\\
		P_{EV,EMS} & t \geq \SI{52}{\second}
	\end{cases}
	\label{eq:delayup}       
\end{equation}

The ramp down time until a lower power set point is reached is shorter with \SI{4}{\second} on average.  We model the ramp down for a requested power decrease is according to Eq. \ref{eq:delaydown} , where $t$ denotes the time after the the new power set point has been set.
\begin{equation}
	P_{ramp_{-}}(t) = 
	\begin{cases}
		P(t=0)  & 0 \leq t \leq \SI{4}{\second}\\
		P_{EV,EMS} & t \geq \SI{4}{\second}
	\end{cases}      
	\label{eq:delaydown} 
\end{equation}
In Fig. \ref{fig:emsvsev} a measurement of a charging process of the EV is shown with the requested charging power by the EMS and the actual charging power of the EV. 

\section{Conclusion}
In this publication we parameterized a comprehensive EV simulation model.
We parameterized a model for the traction battery, charger and charging process control via an EMS system for the Smart e.d. (2013).
The electrical model of the battery cell (50 Ah, manufactured by li-tec (Daimler AG)) is based on an equivalent circuit diagram  (Fig. \ref{fig:eqd-cell}) with a serial resistance, two RC circuits and a voltage source (dual polarization Thevenin model). It was parameterized via capacity tests, OCV measurements, pulse tests and impedance spectroscopy tests. The two time constants $\tau_{1}$ and $\tau_{2}$ of the electrical model are in the range of $\SI{e-2}{\second}$ to $\SI{e-1}{\second}$ and  $\SI{e1}{\second}$ to $\SI{e2}{\second}$ range respectively. 
The electrical model of the battery cell was scaled to obtain the electrical model of the traction battery pack (\SI{17.6}{\KWH}, 93s1p, manufactured by Deutsche Accumotive).  In order to parameterize the thermal model of the traction battery pack we disassembled a battery pack and collected measurements of its size, composition and materials (Fig. \ref{fig:packdistributions}). We parameterized a simplified lumped thermal model (Fig. \ref{fig:eqc:thermal}) using these measurements, literature values of thermal parameters, a liquid cooling system model and laboratory pack measurements.\\ Furthermore, we measured the 1-phase and 3-phase charging efficiency of the on-board charger. The maximum efficiency of 93 \% was reached with a charging power of \SI{11}{\KW}. We also measured the delay between setting a new charging current limitation via the charging cable and the time when the new charging current is reached. On average it took \SI{52}{\second} to reach a higher charging current by the EV and \SI{4}{\second} to reach a lower charging current.\\  We carried out extensive aging tests on the battery cells to measure the aging trends due to calendar and cyclic aging. We observed accelerated calendar aging at high temperatures and SOCs above 75 \%. A linear function of time fit the measurement data of capacity and inner resistance during the aging test. Cells stored at 100 \% SOC and \SI{40}{\degreeCelsius} reached end-of-life (80 \% of initial capacity) after 431 days to 589 days. In the cycle aging test, the first cell to reach end-of-life had been cycled with a DOD of 95 \% around a mean SOC of 50 \%. The end-of-life was reached in the range of 2649 EQFC to 2849 EQFC. An EQFC of 2649 is equivalent to a driving distance of over \SI{306000}{\kilo \meter} for an average consumption of \SI[per-mode=symbol,sticky-per]{15.2}{\kWh \per 100 \kilo \meter}. However, this DOD is not achievable in the Smart e.d. as the BMS limits the SOC range between 3.2 \% and 95.3 \%. The maximum DOD that could be reached is therefore 92.1 \% and would also only be reached in V2G applications as drivers would not take the risk to fully discharge the battery. The second cell to reach EOL is the cell that was cycled with 80 \% DOD around an SOC of 50 \%. The EOL is reached after 3634 EQFC which equates to a driving distance of over \SI{420000}{\kilo \meter}. Higher DODs lead to accelerated aging of the battery cells but overall the impact of cycle aging of the Li-Tec cell of the Smart e.d. is small compared to the impact of calendar aging. This holds especially true if the EV is primarily used for mobility. In V2G applications, such as energy trading, participation in reserve markets or grid boosting the cycle life might play a larger role. In the primary use case of mobility calendar aging is the dominant aging factor for the traction battery pack studied in this paper.\\
The complete traction battery model was validated using laboratory pack measurement tests and measured battery data collected during driving tests via the CAN-Bus of the Smart. The simulation results of the parameterized EV model showed good agreement with the validation data. The RMSE of the cell voltage was between \SI{18.49}{\milli \volt} and \SI{67.17}{\milli \volt} for the laboratory pack and the EV operation tests. The RMSE of the pack temperature was between \SI{1.05}{\kelvin} and \SI{1.99}{\kelvin} for the laboratory pack tests.\\ The full parameter set of the traction battery model is provided in appendix \ref{sec:appendix}. The model presented here is specifically suited to serve as a resource for vehicle-to-grid strategy development as it accurately describes the relevant components of the EV and charger for vehicle-to-grid applications.

\section*{Funding}
The study was conducted within the framework of the GoELK project (research project number 16SBS001C) funded by the German Federal Ministry of Transport and Digital Infrastructure.

\section*{Acknowledgment}
The authors of this publication are solely responsible for its content.

%% file: sections/Tables/C1_table.tex
\begin{tabular}{lllllll}
SOC & \SI{-15}{\degreeCelsius} & \SI{-5}{\degreeCelsius} & \SI{5}{\degreeCelsius} & \SI{15}{\degreeCelsius} & \SI{25}{\degreeCelsius} & \SI{35}{\degreeCelsius} \\ 
\hline 
0 & 6.2036 & 7.9584 & 7.4502 & 13.0975 & 7.3923 & 67.3145 \\ 
5 & 6.7057 & 7.9584 & 7.0831 & 12.0634 & 7.6907 & 52.7914 \\ 
10 & 7.0351 & 7.9058 & 8.1745 & 9.7993 & 9.7993 & 31.2679 \\ 
15 & 7.0362 & 7.7804 & 8.2376 & 8.5921 & 8.5921 & 20.3526 \\ 
20 & 7.025 & 7.6302 & 7.8927 & 7.8728 & 7.8728 & 15.4208 \\ 
25 & 7.0074 & 7.5035 & 7.6377 & 7.4329 & 7.4329 & 12.8902 \\ 
30 & 6.9953 & 7.3938 & 7.4458 & 7.1292 & 7.1292 & 11.4807 \\ 
35 & 6.9859 & 7.321 & 7.3006 & 6.9232 & 6.9232 & 10.5764 \\ 
40 & 6.9818 & 7.2501 & 7.1929 & 6.766 & 6.766 & 9.9806 \\ 
45 & 6.9876 & 7.2037 & 7.1078 & 6.647 & 6.647 & 9.6004 \\ 
50 & 6.9984 & 7.1727 & 7.0403 & 6.5529 & 6.5529 & 9.2664 \\ 
55 & 7.0174 & 7.1526 & 6.9983 & 6.4947 & 6.4947 & 9.1936 \\ 
60 & 7.0358 & 7.1427 & 6.9704 & 6.4468 & 6.4468 & 8.8458 \\ 
65 & 7.0524 & 7.1361 & 6.9536 & 6.4166 & 6.4166 & 8.6808 \\ 
70 & 7.0797 & 7.1536 & 6.9604 & 6.4035 & 6.4035 & 11.7327 \\ 
75 & 7.1131 & 7.1699 & 6.9866 & 6.4153 & 6.4153 & 8.769 \\ 
80 & 7.1384 & 7.2062 & 7.0192 & 6.44 & 6.44 & 8.891 \\ 
85 & 7.1669 & 7.2432 & 7.0495 & 6.4431 & 6.4431 & 8.9094 \\ 
90 & 7.2004 & 7.272 & 7.0808 & 6.4539 & 6.4539 & 8.9672 \\ 
95 & 7.2463 & 7.3147 & 7.1279 & 6.4731 & 6.4731 & 9.121 \\ 
100 & 7.2979 & 7.4146 & 7.2229 & 6.5136 & 8.3874 & 9.2603 \\ 
\hline 
\end{tabular}

%% file: sections/Tables/C2_table.tex
\begin{tabular}{lllllll}
SOC & \SI{-15}{\degreeCelsius} & \SI{-5}{\degreeCelsius} & \SI{5}{\degreeCelsius} & \SI{15}{\degreeCelsius} & \SI{25}{\degreeCelsius} & \SI{35}{\degreeCelsius} \\ 
\hline 
0 & 4141.5919 & 4286.6052 & 5516.4437 & 2118.5361 & 60.0993 & 5531.9622 \\ 
5 & 5022.3094 & 4286.6052 & 6159.9236 & 3033.2749 & 58.1937 & 6678.5867 \\ 
10 & 5708.5295 & 4697.046 & 7798.3847 & 4918.0826 & 4918.0826 & 9448.3505 \\ 
15 & 6041.7281 & 5284.4166 & 8813.2291 & 6317.1787 & 6317.1787 & 11285.6717 \\ 
20 & 6374.0387 & 5884.8955 & 9432.6452 & 7257.1822 & 7257.1822 & 12596.6965 \\ 
25 & 6708.384 & 6360.1984 & 9963.4325 & 7980.8923 & 7980.8923 & 13341.3285 \\ 
30 & 6518.6387 & 6706.3331 & 10449.2833 & 8444.132 & 8444.1321 & 13997.3891 \\ 
35 & 6397.0799 & 7245.0718 & 10766.5061 & 8976.0607 & 8976.0606 & 14755.9898 \\ 
40 & 6431.3843 & 7358.5873 & 10865.4071 & 9414.1746 & 9414.1746 & 15398.4024 \\ 
45 & 6525.6152 & 7433.5444 & 10937.0734 & 9504.4511 & 9504.451 & 16183.0849 \\ 
50 & 6657.4635 & 7507.8527 & 11099.9621 & 9544.943 & 9544.943 & 16387.729 \\ 
55 & 6923.1851 & 7640.8721 & 11096.09 & 9674.996 & 9674.996 & 17946.6885 \\ 
60 & 7163.0191 & 7773.424 & 11214.6962 & 9685.1701 & 9685.1701 & 16451.9069 \\ 
65 & 7360.0617 & 7546.9592 & 11164.4429 & 9733.1104 & 9733.1105 & 16327.2941 \\ 
70 & 7291.066 & 7660.2968 & 11049.7543 & 9769.1101 & 9769.1101 & 19108.3983 \\ 
75 & 7107.4896 & 7490.5601 & 10869.0289 & 9647.1047 & 9647.1047 & 18348.4586 \\ 
80 & 7378.1886 & 7369.5919 & 11006.2558 & 9745.2247 & 9745.2247 & 16253.7877 \\ 
85 & 7507.6149 & 7374.4778 & 11587.7219 & 10077.6235 & 10077.6234 & 16871.9588 \\ 
90 & 7317.6989 & 7519.7098 & 12025.6155 & 10226.4339 & 10226.4339 & 17445.8639 \\ 
95 & 7615.8306 & 7787.3085 & 12402.6332 & 10509.1678 & 10509.1677 & 17726.0212 \\ 
100 & 8229.7746 & 8162.5097 & 12688.7936 & 10981.5609 & 63.7065 & 18549.9549 \\ 
\hline 
\end{tabular}

%% file: sections/Tables/R1_table.tex
\begin{tabular}{lllllll}
SOC & \SI{-15}{\degreeCelsius} & \SI{-5}{\degreeCelsius} & \SI{5}{\degreeCelsius} & \SI{15}{\degreeCelsius} & \SI{25}{\degreeCelsius} & \SI{35}{\degreeCelsius} \\ 
\hline 
0 & 0.080049 & 0.022681 & 0.0075994 & 0.0031085 & 0.0016543 & 0.00083272 \\ 
5 & 0.080813 & 0.022681 & 0.0074289 & 0.0029965 & 0.0016361 & 0.00072845 \\ 
10 & 0.081019 & 0.022443 & 0.0072283 & 0.0027771 & 0.0027771 & 0.00055252 \\ 
15 & 0.080114 & 0.02204 & 0.0069696 & 0.0026011 & 0.0026011 & 0.00047537 \\ 
20 & 0.079403 & 0.02157 & 0.0067263 & 0.0024714 & 0.0024714 & 0.00043031 \\ 
25 & 0.07879 & 0.021096 & 0.006519 & 0.0023783 & 0.0023783 & 0.00040173 \\ 
30 & 0.077989 & 0.020618 & 0.0063371 & 0.0022976 & 0.0022976 & 0.00037995 \\ 
35 & 0.07717 & 0.020242 & 0.0061748 & 0.0022282 & 0.0022282 & 0.00036301 \\ 
40 & 0.076315 & 0.019882 & 0.0060269 & 0.0021654 & 0.0021654 & 0.00034835 \\ 
45 & 0.075792 & 0.01956 & 0.0059057 & 0.0021135 & 0.0021135 & 0.00033465 \\ 
50 & 0.075435 & 0.019288 & 0.0058104 & 0.0020704 & 0.0020704 & 0.00032324 \\ 
55 & 0.074636 & 0.019064 & 0.0057405 & 0.0020316 & 0.0020316 & 0.00031829 \\ 
60 & 0.073822 & 0.01886 & 0.0056725 & 0.0019981 & 0.0019981 & 0.00030497 \\ 
65 & 0.072979 & 0.018613 & 0.005597 & 0.0019707 & 0.0019707 & 0.00029527 \\ 
70 & 0.071834 & 0.018422 & 0.0055279 & 0.0019368 & 0.0019368 & 0.00040059 \\ 
75 & 0.070536 & 0.018181 & 0.0054476 & 0.0019037 & 0.0019037 & 0.00028212 \\ 
80 & 0.068815 & 0.017865 & 0.0053462 & 0.0018703 & 0.0018703 & 0.00027528 \\ 
85 & 0.067044 & 0.01751 & 0.0052494 & 0.0018425 & 0.0018425 & 0.00026975 \\ 
90 & 0.065175 & 0.017177 & 0.0051842 & 0.0018259 & 0.0018259 & 0.00026606 \\ 
95 & 0.064342 & 0.01697 & 0.0051215 & 0.0018122 & 0.0018122 & 0.00026224 \\ 
100 & 0.063995 & 0.017087 & 0.005135 & 0.0017995 & 0.00025511 & 0.00026057 \\ 
\hline 
\end{tabular}

%% file: sections/Tables/R2_table.tex
\begin{tabular}{lllllll}
SOC & \SI{-15}{\degreeCelsius} & \SI{-5}{\degreeCelsius} & \SI{5}{\degreeCelsius} & \SI{15}{\degreeCelsius} & \SI{25}{\degreeCelsius} & \SI{35}{\degreeCelsius} \\ 
\hline 
0 & 0.017962 & 0.0064332 & 0.0053752 & 0.0022864 & 0.0024531 & 0.0013569 \\ 
5 & 0.018272 & 0.0064332 & 0.0050855 & 0.0018491 & 0.0020993 & 0.0010403 \\ 
10 & 0.018677 & 0.0064312 & 0.0051359 & 0.0016798 & 0.0016798 & 0.00078285 \\ 
15 & 0.019272 & 0.0063328 & 0.0051021 & 0.001635 & 0.001635 & 0.00067566 \\ 
20 & 0.019571 & 0.0062175 & 0.0050268 & 0.0016498 & 0.0016498 & 0.00064434 \\ 
25 & 0.019723 & 0.0061373 & 0.0049072 & 0.0016547 & 0.0016547 & 0.00062327 \\ 
30 & 0.018976 & 0.0060186 & 0.0047806 & 0.0016445 & 0.0016445 & 0.0006072 \\ 
35 & 0.018648 & 0.0060566 & 0.0047452 & 0.0016123 & 0.0016123 & 0.00059305 \\ 
40 & 0.019136 & 0.0059892 & 0.0047685 & 0.0015945 & 0.0015945 & 0.00058318 \\ 
45 & 0.01889 & 0.0060495 & 0.0047247 & 0.0016043 & 0.0016043 & 0.00058044 \\ 
50 & 0.018277 & 0.0060949 & 0.0047132 & 0.0016106 & 0.0016106 & 0.00058404 \\ 
55 & 0.01845 & 0.0061307 & 0.0047856 & 0.0016312 & 0.0016312 & 0.00066865 \\ 
60 & 0.018579 & 0.0061718 & 0.0048406 & 0.0016228 & 0.0016228 & 0.00060349 \\ 
65 & 0.018618 & 0.005976 & 0.0048874 & 0.0016475 & 0.0016475 & 0.00061131 \\ 
70 & 0.019384 & 0.0059645 & 0.00508 & 0.0016698 & 0.0016698 & 0.00083064 \\ 
75 & 0.020515 & 0.0061409 & 0.0052437 & 0.001762 & 0.001762 & 0.0008464 \\ 
80 & 0.019859 & 0.0062924 & 0.0051007 & 0.001704 & 0.001704 & 0.0007167 \\ 
85 & 0.01902 & 0.0060444 & 0.0047286 & 0.0016229 & 0.0016229 & 0.00069362 \\ 
90 & 0.017825 & 0.0055591 & 0.0044844 & 0.0015623 & 0.0015623 & 0.00068559 \\ 
95 & 0.016545 & 0.0053424 & 0.0043247 & 0.0015262 & 0.0015262 & 0.00068296 \\ 
100 & 0.015225 & 0.0050156 & 0.0043414 & 0.0014648 & 0.00043849 & 0.00070003 \\ 
\hline 
\end{tabular}

%% file: sections/Tables/Rser_table.tex
\begin{tabular}{lllllll}
SOC & \SI{-15}{\degreeCelsius} & \SI{-5}{\degreeCelsius} & \SI{5}{\degreeCelsius} & \SI{15}{\degreeCelsius} & \SI{25}{\degreeCelsius} & \SI{35}{\degreeCelsius} \\ 
\hline 
0 & 0.0023895 & 0.0018717 & 0.0011043 & 0.00092052 & 0.00072483 & 0.00066083 \\ 
5 & 0.002708 & 0.0018717 & 0.0010703 & 0.00089389 & 0.00073228 & 0.00063753 \\ 
10 & 0.0029141 & 0.0018574 & 0.0011518 & 0.00083587 & 0.00083587 & 0.00059699 \\ 
15 & 0.0028961 & 0.0018258 & 0.0011529 & 0.00079821 & 0.00079821 & 0.00056968 \\ 
20 & 0.0028806 & 0.0017895 & 0.0011236 & 0.00077344 & 0.00077344 & 0.00055148 \\ 
25 & 0.0028664 & 0.0017592 & 0.0011007 & 0.00075798 & 0.00075798 & 0.00053972 \\ 
30 & 0.0028439 & 0.0017321 & 0.0010823 & 0.00074604 & 0.00074604 & 0.00053164 \\ 
35 & 0.0028218 & 0.0017115 & 0.0010676 & 0.00073729 & 0.00073729 & 0.00052591 \\ 
40 & 0.0028001 & 0.0016923 & 0.0010559 & 0.00072961 & 0.00072961 & 0.00052104 \\ 
45 & 0.0027863 & 0.0016799 & 0.0010462 & 0.00072353 & 0.00072353 & 0.00051694 \\ 
50 & 0.0027766 & 0.0016695 & 0.0010386 & 0.0007186 & 0.0007186 & 0.00051324 \\ 
55 & 0.0027659 & 0.0016603 & 0.0010338 & 0.00071487 & 0.00071487 & 0.00051207 \\ 
60 & 0.0027513 & 0.001653 & 0.0010291 & 0.00071147 & 0.00071147 & 0.00050736 \\ 
65 & 0.0027293 & 0.001643 & 0.0010244 & 0.00070907 & 0.00070907 & 0.00050419 \\ 
70 & 0.0027054 & 0.001637 & 0.0010215 & 0.0007063 & 0.0007063 & 0.0005252 \\ 
75 & 0.0026805 & 0.0016297 & 0.0010183 & 0.00070426 & 0.00070426 & 0.00049973 \\ 
80 & 0.0026493 & 0.0016201 & 0.0010142 & 0.00070208 & 0.00070208 & 0.00049713 \\ 
85 & 0.0026158 & 0.0016098 & 0.0010108 & 0.00070045 & 0.00070045 & 0.00049454 \\ 
90 & 0.0025779 & 0.0016007 & 0.0010086 & 0.00069938 & 0.00069938 & 0.00049214 \\ 
95 & 0.0025621 & 0.0015961 & 0.0010072 & 0.00069868 & 0.00069868 & 0.00048918 \\ 
100 & 0.0025566 & 0.0016054 & 0.0010105 & 0.0006979 & 0.00054062 & 0.00048606 \\ 
\hline 
\end{tabular}

%% file: sections/Tables/OCV_table.tex
\begin{tabular}{llllll}
SOC & \SI{-5}{\degreeCelsius} & \SI{5}{\degreeCelsius} & \SI{15}{\degreeCelsius} & \SI{25}{\degreeCelsius} & \SI{35}{\degreeCelsius} \\ 
\hline 
-5 & 3.3785 & 3.4228 & 3.4152 & 3.2082 & 3.1035 \\ 
0 & 3.4064 & 3.4481 & 3.4427 & 3.3287 & 3.3076 \\ 
5 & 3.4342 & 3.4734 & 3.4703 & 3.4477 & 3.4526 \\ 
10 & 3.4621 & 3.4987 & 3.4978 & 3.4963 & 3.498 \\ 
15 & 3.49 & 3.524 & 3.5254 & 3.5223 & 3.5229 \\ 
20 & 3.5178 & 3.5494 & 3.5523 & 3.5498 & 3.5499 \\ 
25 & 3.5457 & 3.5746 & 3.577 & 3.5754 & 3.5748 \\ 
30 & 3.5735 & 3.5965 & 3.5991 & 3.5974 & 3.5967 \\ 
35 & 3.6014 & 3.6194 & 3.6234 & 3.6214 & 3.6204 \\ 
40 & 3.6292 & 3.6427 & 3.6485 & 3.6498 & 3.6496 \\ 
45 & 3.6571 & 3.6648 & 3.6702 & 3.6715 & 3.6749 \\ 
50 & 3.685 & 3.6884 & 3.6927 & 3.6936 & 3.6978 \\ 
55 & 3.7128 & 3.715 & 3.7192 & 3.7201 & 3.7241 \\ 
60 & 3.7451 & 3.7469 & 3.7509 & 3.7516 & 3.7553 \\ 
65 & 3.7825 & 3.7839 & 3.7887 & 3.7892 & 3.7922 \\ 
70 & 3.8254 & 3.8274 & 3.8313 & 3.8316 & 3.8341 \\ 
75 & 3.8746 & 3.8751 & 3.8789 & 3.8794 & 3.8805 \\ 
80 & 3.9326 & 3.9303 & 3.934 & 3.9357 & 3.9342 \\ 
85 & 3.9954 & 3.996 & 3.9978 & 3.998 & 3.9966 \\ 
90 & 4.0589 & 4.0586 & 4.0584 & 4.0574 & 4.0564 \\ 
95 & 4.1242 & 4.1209 & 4.1198 & 4.1176 & 4.1165 \\ 
100 & 4.1862 & 4.1862 & 4.1862 & 4.1835 & 4.1816 \\ 
105 & 4.1862 & 4.1862 & 4.1862 & 4.1862 & 4.1862 \\ 
\hline 
\end{tabular}

%% file: SmartFlex-EV-Model.bbl
\begin{thebibliography}{10}
\expandafter\ifx\csname url\endcsname\relax
  \def\url#1{\texttt{#1}}\fi
\expandafter\ifx\csname urlprefix\endcsname\relax\def\urlprefix{URL }\fi
\expandafter\ifx\csname href\endcsname\relax
  \def\href#1#2{#2} \def\path#1{#1}\fi

\bibitem{EuropeanStrategy}
{European Comission}, \href{https://ec.europa.eu/clima/policies/transport_en}{A
  european strategy for low-emission mobility, {SWD(2016) 244 final}} 2016.
\newline\urlprefix\url{https://ec.europa.eu/clima/policies/transport_en}

\bibitem{eurostat}
t.~S. O. o. t. E.~U. European Commission~Eurostat, Energy dashboard,
  \url{https://ec.europa.eu/eurostat/cache/infographs/energy_dashboard/endash.html}.

\bibitem{ieaev}
IEA, Global ev outlook 2020, Tech. rep., International Energy Agency (2020).

\bibitem{infasInstitutfurangewandteSozialwissenschaftGmbH.2017}
{infas Institut f{\"u}r angewandte Sozialwissenschaft GmbH}, Mobilit{\"a}t in
  deutschland 2017 - ergebnisbericht (2017).

\bibitem{MESBAHI2021102260}
T.~Mesbahi, R.~B. Sugrañes, R.~Bakri, P.~Bartholomeüs,
  \href{https://www.sciencedirect.com/science/article/pii/S2352152X21000281}{Coupled
  electro-thermal modeling of lithium-ion batteries for electric vehicle
  application}, Journal of Energy Storage 35 (2021) 102260.
\newblock \href {https://doi.org/https://doi.org/10.1016/j.est.2021.102260}
  {\path{doi:https://doi.org/10.1016/j.est.2021.102260}}.
\newline\urlprefix\url{https://www.sciencedirect.com/science/article/pii/S2352152X21000281}

\bibitem{SCHMID2020101736}
M.~Schmid, U.~Vögele, C.~Endisch,
  \href{https://www.sciencedirect.com/science/article/pii/S2352152X20315735}{A
  novel matrix-vector-based framework for modeling and simulation of electric
  vehicle battery packs}, Journal of Energy Storage 32 (2020) 101736.
\newblock \href {https://doi.org/https://doi.org/10.1016/j.est.2020.101736}
  {\path{doi:https://doi.org/10.1016/j.est.2020.101736}}.
\newline\urlprefix\url{https://www.sciencedirect.com/science/article/pii/S2352152X20315735}

\bibitem{ZHU2019113339}
R.~Zhu, B.~Duan, C.~Zhang, S.~Gong,
  \href{https://www.sciencedirect.com/science/article/pii/S030626191931013X}{Accurate
  lithium-ion battery modeling with inverse repeat binary sequence for electric
  vehicle applications}, Applied Energy 251 (2019) 113339.
\newblock \href
  {https://doi.org/https://doi.org/10.1016/j.apenergy.2019.113339}
  {\path{doi:https://doi.org/10.1016/j.apenergy.2019.113339}}.
\newline\urlprefix\url{https://www.sciencedirect.com/science/article/pii/S030626191931013X}

\bibitem{electronics8080834}
F.~Wen, B.~Duan, C.~Zhang, R.~Zhu, Y.~Shang, J.~Zhang,
  \href{https://www.mdpi.com/2079-9292/8/8/834}{High-accuracy parameter
  identification method for equivalent-circuit models of lithium-ion batteries
  based on the stochastic theory response reconstruction}, Electronics 8~(8)
  (2019).
\newblock \href {https://doi.org/10.3390/electronics8080834}
  {\path{doi:10.3390/electronics8080834}}.
\newline\urlprefix\url{https://www.mdpi.com/2079-9292/8/8/834}

\bibitem{8892901}
C.~Irimia, M.~Grovu, G.-M. Sirbu, A.~Birtas, C.~Husar, M.~Ponchant, The
  modeling and simulation of an electric vehicle based on simcenter amesim
  platform, in: 2019 Electric Vehicles International Conference (EV), 2019, pp.
  1--6.
\newblock \href {https://doi.org/10.1109/EV.2019.8892901}
  {\path{doi:10.1109/EV.2019.8892901}}.

\bibitem{HOSSEINZADEH201877}
E.~Hosseinzadeh, R.~Genieser, D.~Worwood, A.~Barai, J.~Marco, P.~Jennings,
  \href{https://www.sciencedirect.com/science/article/pii/S0378775318301411}{A
  systematic approach for electrochemical-thermal modelling of a large format
  lithium-ion battery for electric vehicle application}, Journal of Power
  Sources 382 (2018) 77--94.
\newblock \href
  {https://doi.org/https://doi.org/10.1016/j.jpowsour.2018.02.027}
  {\path{doi:https://doi.org/10.1016/j.jpowsour.2018.02.027}}.
\newline\urlprefix\url{https://www.sciencedirect.com/science/article/pii/S0378775318301411}

\bibitem{8070984}
M.~Jafari, A.~Gauchia, S.~Zhao, K.~Zhang, L.~Gauchia, Electric vehicle battery
  cycle aging evaluation in real-world daily driving and vehicle-to-grid
  services, IEEE Transactions on Transportation Electrification 4~(1) (2018)
  122--134.
\newblock \href {https://doi.org/10.1109/TTE.2017.2764320}
  {\path{doi:10.1109/TTE.2017.2764320}}.

\bibitem{GAO2017103}
Y.~Gao, J.~Jiang, C.~Zhang, W.~Zhang, Z.~Ma, Y.~Jiang,
  \href{https://www.sciencedirect.com/science/article/pii/S0378775317305876}{Lithium-ion
  battery aging mechanisms and life model under different charging stresses},
  Journal of Power Sources 356 (2017) 103--114.
\newblock \href
  {https://doi.org/https://doi.org/10.1016/j.jpowsour.2017.04.084}
  {\path{doi:https://doi.org/10.1016/j.jpowsour.2017.04.084}}.
\newline\urlprefix\url{https://www.sciencedirect.com/science/article/pii/S0378775317305876}

\bibitem{7006731}
J.~Jaguemont, L.~Boulon, Y.~Dubé, Characterization and modeling of a
  hybrid-electric-vehicle lithium-ion battery pack at low temperatures, IEEE
  Transactions on Vehicular Technology 65~(1) (2016) 1--14.
\newblock \href {https://doi.org/10.1109/TVT.2015.2391053}
  {\path{doi:10.1109/TVT.2015.2391053}}.

\bibitem{Schmalstieg.2014}
J.~Schmalstieg, S.~K{\"a}bitz, M.~Ecker, D.~U. Sauer, A holistic aging model
  for li(nimnco)o2 based 18650 lithium-ion batteries, Journal of Power Sources
  257 (2014) 325--334.
\newblock \href {https://doi.org/10.1016/j.jpowsour.2014.02.012}
  {\path{doi:10.1016/j.jpowsour.2014.02.012}}.

\bibitem{WANG2020110015}
Y.~Wang, J.~Tian, Z.~Sun, L.~Wang, R.~Xu, M.~Li, Z.~Chen,
  \href{https://www.sciencedirect.com/science/article/pii/S1364032120303063}{A
  comprehensive review of battery modeling and state estimation approaches for
  advanced battery management systems}, Renewable and Sustainable Energy
  Reviews 131 (2020) 110015.
\newblock \href {https://doi.org/https://doi.org/10.1016/j.rser.2020.110015}
  {\path{doi:https://doi.org/10.1016/j.rser.2020.110015}}.
\newline\urlprefix\url{https://www.sciencedirect.com/science/article/pii/S1364032120303063}

\bibitem{Newman1975PorouselectrodeTW}
J.~Newman, W.~Tiedemann, Porous‐electrode theory with battery applications,
  Aiche Journal 21 (1975) 25--41.

\bibitem{osti_142201}
T.~F. Fuller, M.~Doyle, J.~Newman,
  \href{https://www.osti.gov/biblio/142201}{Simulation and optimization of the
  dual lithium ion insertion cell}, Journal of the Electrochemical Society
  141~(1) (1 1994).
\newblock \href {https://doi.org/10.1149/1.2054684}
  {\path{doi:10.1149/1.2054684}}.
\newline\urlprefix\url{https://www.osti.gov/biblio/142201}

\bibitem{electrochemicalsystems}
J.~N. und Karen E. Thomas-Alyea., Electrochemical Systems., 3rd Edition,
  Wiley-Interscience, New York, 2004.

\bibitem{ROMEROBECERRIL201110267}
A.~Romero-Becerril, L.~Alvarez-Icaza,
  \href{https://www.sciencedirect.com/science/article/pii/S0378775311013280}{Comparison
  of discretization methods applied to the single-particle model of lithium-ion
  batteries}, Journal of Power Sources 196~(23) (2011) 10267--10279.
\newblock \href
  {https://doi.org/https://doi.org/10.1016/j.jpowsour.2011.06.091}
  {\path{doi:https://doi.org/10.1016/j.jpowsour.2011.06.091}}.
\newline\urlprefix\url{https://www.sciencedirect.com/science/article/pii/S0378775311013280}

\bibitem{JOHNSON2002321}
V.~Johnson,
  \href{https://www.sciencedirect.com/science/article/pii/S0378775302001945}{Battery
  performance models in advisor}, Journal of Power Sources 110~(2) (2002)
  321--329.
\newblock \href {https://doi.org/https://doi.org/10.1016/S0378-7753(02)00194-5}
  {\path{doi:https://doi.org/10.1016/S0378-7753(02)00194-5}}.
\newline\urlprefix\url{https://www.sciencedirect.com/science/article/pii/S0378775302001945}

\bibitem{6652363}
L.~Mihet-Popa, O.~M.~F. Camacho, P.~B. Nørgård, Charging and discharging
  tests for obtaining an accurate dynamic electro-thermal model of high power
  lithium-ion pack system for hybrid and ev applications, in: 2013 IEEE
  Grenoble Conference, 2013, pp. 1--6.
\newblock \href {https://doi.org/10.1109/PTC.2013.6652363}
  {\path{doi:10.1109/PTC.2013.6652363}}.

\bibitem{Liaw2004835}
B.~Liaw, G.~Nagasubramanian, R.~Jungst, D.~Doughty,
  \href{https://www.scopus.com/inward/record.uri?eid=2-s2.0-10144244017&doi=10.1016%2fj.ssi.2004.09.049&partnerID=40&md5=b286ce2820221ed7a1d776bc4097a2f7}{Modeling
  of lithium ion cells - a simple equivalent-circuit model approach}, Solid
  State Ionics 175~(1-4) (2004) 835--839, cited By 190.
\newblock \href {https://doi.org/10.1016/j.ssi.2004.09.049}
  {\path{doi:10.1016/j.ssi.2004.09.049}}.
\newline\urlprefix\url{https://www.scopus.com/inward/record.uri?eid=2-s2.0-10144244017&doi=10.1016%2fj.ssi.2004.09.049&partnerID=40&md5=b286ce2820221ed7a1d776bc4097a2f7}

\bibitem{6237284}
H.~Rahimi-Eichi, F.~Baronti, M.-Y. Chow, Modeling and online parameter
  identification of li-polymer battery cells for soc estimation, in: 2012 IEEE
  International Symposium on Industrial Electronics, 2012, pp. 1336--1341.
\newblock \href {https://doi.org/10.1109/ISIE.2012.6237284}
  {\path{doi:10.1109/ISIE.2012.6237284}}.

\bibitem{6108373}
A.~H. Ranjbar, A.~Banaei, A.~Khoobroo, B.~Fahimi, Online estimation of state of
  charge in li-ion batteries using impulse response concept, IEEE Transactions
  on Smart Grid 3~(1) (2012) 360--367.
\newblock \href {https://doi.org/10.1109/TSG.2011.2169818}
  {\path{doi:10.1109/TSG.2011.2169818}}.

\bibitem{MINGANT2021102592}
R.~Mingant, M.~Petit, S.~Belaïd, J.~Bernard,
  \href{https://www.sciencedirect.com/science/article/pii/S2352152X21003352}{Data-driven
  model development to predict the aging of a li-ion battery pack in electric
  vehicles representative conditions}, Journal of Energy Storage 39 (2021)
  102592.
\newblock \href {https://doi.org/https://doi.org/10.1016/j.est.2021.102592}
  {\path{doi:https://doi.org/10.1016/j.est.2021.102592}}.
\newline\urlprefix\url{https://www.sciencedirect.com/science/article/pii/S2352152X21003352}

\bibitem{TANG2019113591}
X.~Tang, C.~Zou, K.~Yao, J.~Lu, Y.~Xia, F.~Gao,
  \href{https://www.sciencedirect.com/science/article/pii/S0306261919312656}{Aging
  trajectory prediction for lithium-ion batteries via model migration and
  bayesian monte carlo method}, Applied Energy 254 (2019) 113591.
\newblock \href
  {https://doi.org/https://doi.org/10.1016/j.apenergy.2019.113591}
  {\path{doi:https://doi.org/10.1016/j.apenergy.2019.113591}}.
\newline\urlprefix\url{https://www.sciencedirect.com/science/article/pii/S0306261919312656}

\bibitem{TANG2021100302}
X.~Tang, K.~Liu, K.~Li, W.~D. Widanage, E.~Kendrick, F.~Gao,
  \href{https://www.sciencedirect.com/science/article/pii/S2666389921001458}{Recovering
  large-scale battery aging dataset with machine learning}, Patterns 2~(8)
  (2021) 100302.
\newblock \href {https://doi.org/https://doi.org/10.1016/j.patter.2021.100302}
  {\path{doi:https://doi.org/10.1016/j.patter.2021.100302}}.
\newline\urlprefix\url{https://www.sciencedirect.com/science/article/pii/S2666389921001458}

\bibitem{Carlier2002}
D.~Carlier, I.~Saadoune, M.~M{\'{e}}n{\'{e}}trier, C.~Delmas,
  \href{https://doi.org/10.1149/1.1503075}{Lithium electrochemical
  deintercalation from o2-{LiCoO}[sub 2]}, Journal of The Electrochemical
  Society 149~(10) (2002) A1310.
\newblock \href {https://doi.org/10.1149/1.1503075}
  {\path{doi:10.1149/1.1503075}}.
\newline\urlprefix\url{https://doi.org/10.1149/1.1503075}

\bibitem{TRAN2020101785}
M.-K. Tran, A.~Mevawala, S.~Panchal, K.~Raahemifar, M.~Fowler, R.~Fraser,
  \href{https://www.sciencedirect.com/science/article/pii/S2352152X20316224}{Effect
  of integrating the hysteresis component to the equivalent circuit model of
  lithium-ion battery for dynamic and non-dynamic applications}, Journal of
  Energy Storage 32 (2020) 101785.
\newblock \href {https://doi.org/https://doi.org/10.1016/j.est.2020.101785}
  {\path{doi:https://doi.org/10.1016/j.est.2020.101785}}.
\newline\urlprefix\url{https://www.sciencedirect.com/science/article/pii/S2352152X20316224}

\bibitem{batteries7030051}
M.-K. Tran, A.~DaCosta, A.~Mevawalla, S.~Panchal, M.~Fowler,
  \href{https://www.mdpi.com/2313-0105/7/3/51}{Comparative study of equivalent
  circuit models performance in four common lithium-ion batteries: Lfp, nmc,
  lmo, nca}, Batteries 7~(3) (2021).
\newline\urlprefix\url{https://www.mdpi.com/2313-0105/7/3/51}

\bibitem{Habedank_2018}
J.~B. Habedank, L.~Kraft, A.~Rheinfeld, C.~Krezdorn, A.~Jossen, M.~F. Zaeh,
  \href{https://doi.org/10.1149/2.1181807jes}{Increasing the discharge rate
  capability of lithium-ion cells with laser-structured graphite anodes:
  Modeling and simulation}, Journal of The Electrochemical Society 165~(7)
  (2018) A1563--A1573.
\newblock \href {https://doi.org/10.1149/2.1181807jes}
  {\path{doi:10.1149/2.1181807jes}}.
\newline\urlprefix\url{https://doi.org/10.1149/2.1181807jes}

\bibitem{YANG2020}
\href{https://www.sciencedirect.com/science/article/pii/S2095809920303258}{Extreme
  learning machine-based thermal model for lithium-ion batteries of electric
  vehicles under external short circuit}, Engineering (2020).
\newblock \href {https://doi.org/https://doi.org/10.1016/j.eng.2020.08.015}
  {\path{doi:https://doi.org/10.1016/j.eng.2020.08.015}}.
\newline\urlprefix\url{https://www.sciencedirect.com/science/article/pii/S2095809920303258}

\bibitem{ECKER2012248}
M.~Ecker, J.~B. Gerschler, J.~Vogel, S.~Käbitz, F.~Hust, P.~Dechent, D.~U.
  Sauer,
  \href{https://www.sciencedirect.com/science/article/pii/S0378775312008671}{Development
  of a lifetime prediction model for lithium-ion batteries based on extended
  accelerated aging test data}, Journal of Power Sources 215 (2012) 248--257.
\newblock \href
  {https://doi.org/https://doi.org/10.1016/j.jpowsour.2012.05.012}
  {\path{doi:https://doi.org/10.1016/j.jpowsour.2012.05.012}}.
\newline\urlprefix\url{https://www.sciencedirect.com/science/article/pii/S0378775312008671}

\bibitem{ECKER2014839}
M.~Ecker, N.~Nieto, S.~Käbitz, J.~Schmalstieg, H.~Blanke, A.~Warnecke, D.~U.
  Sauer,
  \href{https://www.sciencedirect.com/science/article/pii/S0378775313016510}{Calendar
  and cycle life study of li(nimnco)o2-based 18650 lithium-ion batteries},
  Journal of Power Sources 248 (2014) 839--851.
\newblock \href
  {https://doi.org/https://doi.org/10.1016/j.jpowsour.2013.09.143}
  {\path{doi:https://doi.org/10.1016/j.jpowsour.2013.09.143}}.
\newline\urlprefix\url{https://www.sciencedirect.com/science/article/pii/S0378775313016510}

\bibitem{LEWERENZ201757}
M.~Lewerenz, A.~Marongiu, A.~Warnecke, D.~U. Sauer,
  \href{https://www.sciencedirect.com/science/article/pii/S0378775317312788}{Differential
  voltage analysis as a tool for analyzing inhomogeneous aging: A case study
  for lifepo4|graphite cylindrical cells}, Journal of Power Sources 368 (2017)
  57--67.
\newblock \href
  {https://doi.org/https://doi.org/10.1016/j.jpowsour.2017.09.059}
  {\path{doi:https://doi.org/10.1016/j.jpowsour.2017.09.059}}.
\newline\urlprefix\url{https://www.sciencedirect.com/science/article/pii/S0378775317312788}

\bibitem{Renganathan_2010}
S.~Renganathan, G.~Sikha, S.~Santhanagopalan, R.~E. White,
  \href{https://doi.org/10.1149/1.3261809}{Theoretical analysis of stresses in
  a lithium ion cell}, Journal of The Electrochemical Society 157~(2) (2010)
  A155.
\newblock \href {https://doi.org/10.1149/1.3261809}
  {\path{doi:10.1149/1.3261809}}.
\newline\urlprefix\url{https://doi.org/10.1149/1.3261809}

\bibitem{LARESGOITI2015112}
I.~Laresgoiti, S.~Käbitz, M.~Ecker, D.~U. Sauer,
  \href{https://www.sciencedirect.com/science/article/pii/S0378775315302949}{Modeling
  mechanical degradation in lithium ion batteries during cycling: Solid
  electrolyte interphase fracture}, Journal of Power Sources 300 (2015)
  112--122.
\newblock \href
  {https://doi.org/https://doi.org/10.1016/j.jpowsour.2015.09.033}
  {\path{doi:https://doi.org/10.1016/j.jpowsour.2015.09.033}}.
\newline\urlprefix\url{https://www.sciencedirect.com/science/article/pii/S0378775315302949}

\bibitem{osti_20001062}
P.~Arora, M.~Doyle, R.~E. White,
  \href{https://www.osti.gov/biblio/20001062}{Mathematical modeling of the
  lithium deposition overcharge reaction in lithium-ion batteries using
  carbon-based negative electrodes}, Journal of the Electrochemical Society
  146~(10) (10 1999).
\newblock \href {https://doi.org/10.1149/1.1392512}
  {\path{doi:10.1149/1.1392512}}.
\newline\urlprefix\url{https://www.osti.gov/biblio/20001062}

\bibitem{INLvehicletesting}
A.~G. J.~Diaz, \href{https://www.osti.gov/servlets/purl/1481912}{Advanced
  vehicle testing and evaluation}, Tech. rep., intertek (2018).
\newline\urlprefix\url{https://www.osti.gov/servlets/purl/1481912}

\bibitem{7556282}
D.-H. Kim, M.-J. Kim, B.-K. Lee, An integrated battery charger with high power
  density and efficiency for electric vehicles, IEEE Transactions on Power
  Electronics 32~(6) (2017) 4553--4565.
\newblock \href {https://doi.org/10.1109/TPEL.2016.2604404}
  {\path{doi:10.1109/TPEL.2016.2604404}}.

\bibitem{radimov2020three}
N.~Radimov, G.~Li, M.~Tang, X.~Wang, Three-stage sic-based bi-directional
  on-board battery charger with titanium level efficiency, IET Power
  Electronics 13~(7) (2020) 1477--1480.

\bibitem{9203459}
W.~Schram, N.~Brinkel, G.~Smink, T.~van Wijk, W.~van Sark, Empirical evaluation
  of v2g round-trip efficiency, in: 2020 International Conference on Smart
  Energy Systems and Technologies (SEST), 2020, pp. 1--6.
\newblock \href {https://doi.org/10.1109/SEST48500.2020.9203459}
  {\path{doi:10.1109/SEST48500.2020.9203459}}.

\bibitem{PeterBachAndersenSeyedmostafaHashemiToghroljerdiThomasMeierSrensenBjrnEskeChristense.2019}
{Peter Bach Andersen, Seyedmostafa Hashemi Toghroljerdi, Thomas Meier
  S{\"o}rensen, Bj{\"o}rn Eske Christensen, Jens Christian Morell Lodberg
  H{\"o}j,Antonio Zecchino}, The parker project (2019).

\bibitem{INEES}
G.~Arnold, R.~Brandl, T.~Degner, N.~Gerhardt, M.~Landau, D.~Nestle, M.~Portula,
  A.~Scheidler, R.~Schwinn, K.~Baumbusch, T.~Dörschlag, A.and~Eberhardt,
  V.~Wacker, A.~Wesemann, O.~Führer, T.~Leifert, G.~Bäuml, G.~Bärwaldt,
  H.~Haupt, M.~Kammerlocher, H.~Nannen,
  \href{https://www.erneuerbar-mobil.de/sites/default/files/2016-09/INEES_Abschlussbericht.pdf}{Intelligente
  netzanbindung von elektrofahrzeugen zur erbringung von systemdienstleistungen
  – inees}, Tech. rep., Fraunhofer IWES, LichtBlick SE, SMA AG, Volkswagen AG
  (2018).
\newline\urlprefix\url{https://www.erneuerbar-mobil.de/sites/default/files/2016-09/INEES_Abschlussbericht.pdf}

\bibitem{degner2017grid}
T.~Degner, G.~Arnold, R.~Brandl, J.~Dollichon, A.~Scheidler, Grid impact of
  electric vehicles with secondary control reserve capability, in: Proceeding
  of the 1st E-Mobility Power System Integration Symposium, 2017.

\bibitem{DaimlerAG}
{Daimler AG}, \url{https://media.daimler.com/marsMediaSite/ko/en/9920260},
  accessed on 20.04.2020.

\bibitem{introductionsmart}
{Daimler AG}, Introduction of the smart fortwo electric drive (3rd generation)
  model series 451: Introduction into service manual, Tech. rep.

\bibitem{Witzenhausen:687819}
H.~Witzenhausen,
  \href{https://publications.rwth-aachen.de/record/687819}{{E}lektrische
  {B}atteriespeichermodelle : {M}odellbildung, {P}arameteridentifikation und
  {M}odellreduktion; 1. {A}uflage}, Dissertation, RWTH Aachen University,
  Aachen, veröffentlicht auf dem Publikationsserver der RWTH Aachen
  University; Dissertation, RWTH Aachen University, 2017 (2017).
\newblock \href {https://doi.org/10.18154/RWTH-2017-03437}
  {\path{doi:10.18154/RWTH-2017-03437}}.
\newline\urlprefix\url{https://publications.rwth-aachen.de/record/687819}

\bibitem{LiTecBatteryGmbH.2015}
{Li-Tec Battery GmbH}, Ms-td-058 technisches datenblatt hea 50; rev. 2.0, Tech.
  rep. (2015).

\bibitem{Abada.2016}
S.~Abada, G.~Marlair, A.~Lecocq, M.~Petit, V.~Sauvant-Moynot, F.~Huet, Safety
  focused modeling of lithium-ion batteries: A review, Journal of Power Sources
  306 (2016) 178--192.
\newblock \href {https://doi.org/10.1016/j.jpowsour.2015.11.100}
  {\path{doi:10.1016/j.jpowsour.2015.11.100}}.

\bibitem{JanPhilippSchmidt.2013}
J.~P. Schmidt, Verfahren zur charakterisierung und modellierung von
  lithium-ionen zellen, Ph.D. thesis (2013).
\newblock \href {https://doi.org/10.5445/KSP/1000036622}
  {\path{doi:10.5445/KSP/1000036622}}.

\bibitem{Hu.2012}
X.~Hu, S.~Li, H.~Peng, A comparative study of equivalent circuit models for
  li-ion batteries, Journal of Power Sources 198 (2012) 359--367.
\newblock \href {https://doi.org/10.1016/j.jpowsour.2011.10.013}
  {\path{doi:10.1016/j.jpowsour.2011.10.013}}.

\bibitem{Schmalstieg.2017}
J.~Schmalstieg,
  \href{https://publications.rwth-aachen.de/record/689927}{{P}hysikalisch-elektrochemische
  {S}imulation von {L}ithium-{I}onen-{B}atterien : {I}mplementierung,
  {P}arametrierung und {A}nwendung}, Dissertation, RWTH Aachen University,
  Aachen, veröffentlicht auf dem Publikationsserver der RWTH Aachen
  University; Dissertation, RWTH Aachen University, 2017 (2017).
\newblock \href {https://doi.org/10.18154/RWTH-2017-04693}
  {\path{doi:10.18154/RWTH-2017-04693}}.
\newline\urlprefix\url{https://publications.rwth-aachen.de/record/689927}

\bibitem{8398150}
Q.-Z. Zhang, X.-Y. Wang, H.-M. Yuan, Estimation for soc of li-ion battery based
  on two-order rc temperature model, in: 2018 13th IEEE Conference on
  Industrial Electronics and Applications (ICIEA), 2018, pp. 2601--2606.
\newblock \href {https://doi.org/10.1109/ICIEA.2018.8398150}
  {\path{doi:10.1109/ICIEA.2018.8398150}}.

\bibitem{9236517}
M.~S. Kumar, T.~R. Manasa, B.~Raja, K.~Selvajyothi, Estimation of state of
  charge and terminal voltage of li-ion battery using extended kalman filter,
  in: 2020 6th IEEE International Energy Conference (ENERGYCon), 2020, pp.
  515--520.
\newblock \href {https://doi.org/10.1109/ENERGYCon48941.2020.9236517}
  {\path{doi:10.1109/ENERGYCon48941.2020.9236517}}.

\bibitem{Kiel:228547}
M.~Kiel,
  \href{https://publications.rwth-aachen.de/record/228547}{{I}mpedanzspektroskopie
  an {B}atterien unter besonderer {B}er{\"u}cksichtigung von {B}atteriesensoren
  f{\"u}r den {F}eldeinsatz}, Ph.D. thesis, Aachen, zugl.: Aachen, Techn.
  Hochsch., Diss., 2013 (2013).
\newline\urlprefix\url{https://publications.rwth-aachen.de/record/228547}

\bibitem{Bernardi_1985}
D.~Bernardi, E.~Pawlikowski, J.~Newman,
  \href{https://doi.org/10.1149/1.2113792}{A general energy balance for battery
  systems}, Journal of The Electrochemical Society 132~(1) (1985) 5--12.
\newblock \href {https://doi.org/10.1149/1.2113792}
  {\path{doi:10.1149/1.2113792}}.
\newline\urlprefix\url{https://doi.org/10.1149/1.2113792}

\bibitem{Magnor:696065}
D.~T. Magnor,
  \href{https://publications.rwth-aachen.de/record/696065}{{G}lobale
  {O}ptimierung netzgekoppelter {PV}-{B}atteriesysteme unter besonderer
  {B}er{\"u}cksichtigung der {B}atteriealterung}, Dissertation, RWTH Aachen
  University, Aachen, ver{\"o}ffentlicht auf dem Publikationsserver der RWTH
  Aachen University; Dissertation, RWTH Aachen University, 2017 (2017).
\newblock \href {https://doi.org/10.18154/RWTH-2017-06592}
  {\path{doi:10.18154/RWTH-2017-06592}}.
\newline\urlprefix\url{https://publications.rwth-aachen.de/record/696065}

\bibitem{Hust:752755}
F.~E. Hust,
  \href{https://publications.rwth-aachen.de/record/752755}{{P}hysico-chemically
  motivated parameterization and modelling of real-time capable lithium-ion
  battery models : a case study on the {T}esla {M}odel {S} battery},
  Dissertation, Rheinisch-Westf{\"a}lische Technische Hochschule Aachen,
  Aachen, veröffentlicht auf dem Publikationsserver der RWTH Aachen University
  2019; Dissertation, Rheinisch-Westfälische Technische Hochschule Aachen,
  2018 (2018).
\newblock \href {https://doi.org/10.18154/RWTH-2019-00249}
  {\path{doi:10.18154/RWTH-2019-00249}}.
\newline\urlprefix\url{https://publications.rwth-aachen.de/record/752755}

\bibitem{mobility95}
D.~L. und Raumfahrtzentrum~e.V. (DLR), Mobilitaet in tabellen,
  \url{https://mobilitaet-in-tabellen.dlr.de/mit/}, accessed 24.06.2021 (2017).

\bibitem{AngeEtienneAcquaviva.2012}
{Ange-Etienne Acquaviva}, C{\'e}dric,
  \href{https://docplayer.net/26479902-Diploma-thesis-
  development-and-validation-of-a-plant-model-for-
  battery-monitoring-systems-bms-for-high-voltage-batteries.html}{Development
  and validation of a plant model for battery monitoring systems (bms) for high
  voltage batteries}, Diploma thesis, {INSA Strasbourg} (2012).
\newline\urlprefix\url{https://docplayer.net/26479902-Diploma-thesis-
  development-and-validation-of-a-plant-model-for-
  battery-monitoring-systems-bms-for-high-voltage-batteries.html}

\bibitem{FRIESEN20161}
A.~Friesen, F.~Horsthemke, X.~Mönnighoff, G.~Brunklaus, R.~Krafft, M.~Börner,
  T.~Risthaus, M.~Winter, F.~M. Schappacher,
  \href{https://www.sciencedirect.com/science/article/pii/S0378775316313106}{Impact
  of cycling at low temperatures on the safety behavior of 18650-type lithium
  ion cells: Combined study of mechanical and thermal abuse testing accompanied
  by post-mortem analysis}, Journal of Power Sources 334 (2016) 1--11.
\newblock \href
  {https://doi.org/https://doi.org/10.1016/j.jpowsour.2016.09.120}
  {\path{doi:https://doi.org/10.1016/j.jpowsour.2016.09.120}}.
\newline\urlprefix\url{https://www.sciencedirect.com/science/article/pii/S0378775316313106}

\bibitem{C.Ziebert}
{C. Ziebert, N. Uhlmann, S. Ouyang, B. Lei, W. Zhao, M. Rohde, H.J Seifert},
  Battery calorimetry of li-ion cells to prevent thermal runaway and develop
  safer cells, Tech. rep., Mainz, Germany (2018).

\bibitem{Baehr}
H.~D. Baehr, K.~Stephan, W{\"a}rme- und Stoff{\"u}bertragung, 9. Auflage,
  Springer, Berlin; Heidelberg; New York, 2016.

\bibitem{DOWNIE1980779}
D.~Downie, J.~Martin,
  \href{https://www.sciencedirect.com/science/article/pii/0021961480901767}{An
  adiabatic calorimeter for heat-capacity measurements between 6 and 300 k. the
  molar heat capacity of aluminium}, The Journal of Chemical Thermodynamics
  12~(8) (1980) 779--786.
\newblock \href {https://doi.org/https://doi.org/10.1016/0021-9614(80)90176-7}
  {\path{doi:https://doi.org/10.1016/0021-9614(80)90176-7}}.
\newline\urlprefix\url{https://www.sciencedirect.com/science/article/pii/0021961480901767}

\bibitem{WEIDENFELLER2004423}
B.~Weidenfeller, M.~Höfer, F.~R. Schilling,
  \href{https://www.sciencedirect.com/science/article/pii/S1359835X03003440}{Thermal
  conductivity, thermal diffusivity, and specific heat capacity of particle
  filled polypropylene}, Composites Part A: Applied Science and Manufacturing
  35~(4) (2004) 423--429.
\newblock \href
  {https://doi.org/https://doi.org/10.1016/j.compositesa.2003.11.005}
  {\path{doi:https://doi.org/10.1016/j.compositesa.2003.11.005}}.
\newline\urlprefix\url{https://www.sciencedirect.com/science/article/pii/S1359835X03003440}

\bibitem{BAUMHOFER2014332}
T.~Baumhöfer, M.~Brühl, S.~Rothgang, D.~U. Sauer,
  \href{https://www.sciencedirect.com/science/article/pii/S0378775313014584}{Production
  caused variation in capacity aging trend and correlation to initial cell
  performance}, Journal of Power Sources 247 (2014) 332--338.
\newblock \href
  {https://doi.org/https://doi.org/10.1016/j.jpowsour.2013.08.108}
  {\path{doi:https://doi.org/10.1016/j.jpowsour.2013.08.108}}.
\newline\urlprefix\url{https://www.sciencedirect.com/science/article/pii/S0378775313014584}

\end{thebibliography}
